\documentclass[useAMS,usenatbib]{mnras}
\usepackage[T1]{fontenc}
\usepackage{amsmath}
\usepackage{amssymb}
\usepackage{graphicx}
\usepackage{wasysym}
\usepackage{esint}
\usepackage[authoryear]{natbib}
\usepackage[utf8]{inputenc}
\usepackage{threeparttable}
\usepackage{hyperref}
\usepackage{xspace}
\usepackage{url}
\usepackage{babel}
\usepackage{ulem}
\usepackage{newtxtext,newtxmath}

\newcommand{\arepo}{\textsc{Arepo}\xspace}
\defcitealias{2021WerhahnI}{Paper I}
\defcitealias{2021WerhahnII}{Paper II}
\defcitealias{2021WerhahnIII}{Paper III}

\makeatletter

\pdfpageheight\paperheight
\pdfpagewidth\paperwidth

\usepackage{xcolor}

\title[CRs and non-thermal emission in galaxies III.]{Cosmic rays and non-thermal emission in simulated galaxies: III. probing cosmic ray calorimetry with radio spectra and the FIR-radio correlation}

\author[M. Werhahn et al.]{Maria Werhahn,$^{1,2}$\thanks{E-mail:
mwerhahn@aip.de} Christoph Pfrommer,$^{1}$ Philipp Girichidis$^{1}$ 
\\
$^{1}$Leibniz-Institut f\"ur Astrophysik Potsdam (AIP), An der Sternwarte 16, 14482 Potsdam, Germany\\
$^2$Institut f\"ur Physik und Astronomie, Universit\"at Potsdam, Karl-Liebknecht-Str.\,24/25, 14476 Golm, Germany
}

\makeatother

\begin{document}
\date{Accepted 20XX . Received 20XX}

\maketitle
\pagerange{\pageref{firstpage}--\pageref{lastpage}} \pubyear{2021}

\label{firstpage}
\begin{abstract}
An extinction-free estimator of the star-formation rate (SFR) of galaxies is critical for understanding the high-redshift universe. To this end, the nearly linear, tight correlation of far-infrared (FIR) and radio luminosity of star-forming galaxies is widely used. While the FIR is linked to massive star formation, which also generates shock-accelerated cosmic ray (CR) electrons and radio synchrotron emission, a detailed understanding of the underlying physics is still lacking. Hence, we perform three-dimensional magneto-hydrodynamical (MHD) simulations of isolated galaxies over a broad range of halo masses and SFRs using the moving-mesh code \arepo, and evolve the CR proton energy density self-consistently. In post-processing, we calculate the steady-state spectra of primary, shock-accelerated and secondary CR electrons, which result from hadronic CR proton interactions with the interstellar medium. The resulting total radio luminosities correlate with the FIR luminosities as observed and are dominated by primary CR electrons if we account for anisotropic CR diffusion. The increasing contribution of secondary emission up to 30 per cent in starbursts is compensated by the larger bremsstrahlung and Coulomb losses. CR electrons are in the calorimetric limit and lose most of their energy through inverse Compton interactions with star-light and cosmic microwave background (CMB) photons while less energy is converted to synchrotron emission. This implies steep steady-state synchrotron spectra in starbursts. Interestingly, we find that thermal free-free emission hardens the total radio spectra at high radio frequencies and reconciles calorimetric theory with observations while free-free absorption explains the observed low-frequency flattening towards the central regions of starbursts.
\end{abstract}

\section{Introduction}

The radio emission from star-forming galaxies is attributed to the interaction of CR electrons with an ambient interstellar magnetic field. These highly relativistic charged particles emit synchrotron radiation while gyrating around magnetic field lines, giving rise to a power-law spectrum at radio frequencies. 
Because CRs are accelerated at shocks of supernova (SN) remnants and because the lifetime of radio-emitting CR electrons in galaxies is shorter than tens of millions of years, their existence reveals ongoing star formation in a galaxy. Hence, a correlation between the radio synchrotron emission of star-forming galaxies and tracers of their star formation activity, such as the FIR luminosity, is expected. The latter traces ongoing star formation, because the ultra-violet (UV) light of a young stellar population is absorbed by dust and re-emitted in the FIR. 

Indeed, a tight and nearly linear correlation has been observed between the radio and FIR luminosity of galaxies \citep{1971VanDerKruit,1985Helou,1992Condon,2001Yun,2003Bell,2021Molnar,2021Matthews}. 
Due to the tightness of this relation across different types of star-forming galaxies and across five orders of magnitude in luminosity, the FIR-radio correlation (FRC) is widely used to estimate the SFR of galaxies from their radio luminosities. It has the notable feature of not being affected by dust extinction, which is one of the main uncertainties of other SFR estimators, which employ, e.g., the H$\alpha$ and/or UV emission. This makes the FRC a favourable method for estimating SFRs in particular for high redshifts galaxies, where dust properties are unknown, and where accurate SFR estimators are critical for deciphering the cosmic star formation history. However, it is still unclear, whether the FRC evolves with redshift \citep{2010bLacki,2016Schober}.

A linear FRC naturally emerges if CR electrons calorimetrically lose most of their energy to synchrotron emission \citep{Voelk1989,1996Lisenfeld}. This requires that the corresponding timescale of synchrotron losses is shorter than all other loss processes that apply to CR electrons, i.e., losses due to inverse Compton (IC) and bremsstrahlung emission, as well as Coulomb losses and CR electron escape from the galaxy. The latter process arises due to advection and diffusion, and has been suggested to be relevant in galaxies with low SFRs \citep{2006Thompson,2010Lacki}. But highly star-forming galaxies, such as NGC~253 and M82, are expected to be calorimetric, due to high photon energy densities and magnetic field strengths. However, in this picture, the fully-cooled electron spectra would steepen by unity at high particle energies due to the energy dependence of IC and synchrotron cooling. This would imply steep radio spectra with spectral indices at around 1.1, which is in contrast to observed values of 0.5--0.8. Hence, these hard radio spectra of star-forming galaxies pose a challenge to the calorimetric theory.

To gain insights into these processes, the radio and gamma-ray spectra of famous starburst galaxies such as M82 or NGC~253 have been analyzed with so-called one-zone models in which a galaxy is represented as a single entity and parametrized by average quantities \citep{2004Torres, 2010Lacki,2012Paglione,2013Yoast-Hull,2016Eichmann}. One proposed solution for the problem of too steep radio spectra predicted from calorimetric theory has been bremsstrahlung and ionization losses in starburst galaxies \citep{2006Thompson,2010Lacki}, which could flatten CR electron spectra due to their shallower energy dependence in comparison to IC and synchrotron losses. However, at the same time, these additional processes open new energy loss channels for the radio-emitting CR electrons and thus diminish the resulting radio synchrotron luminosity. This would imply a sublinear FRC at high SFRs, which is in conflict with observations. To cure this problem, a second radio emission process has been proposed that fills in this missing radio emission in form of an increasing contribution of synchrotron emission from secondary electrons that are generated in hadronic CR proton interactions with the ambient interstellar medium \citep[ISM,][]{2006Thompson,2010Lacki}, an efficient process in the dense centers of starbursts. This would successfully maintain the linear FRC, even at high SFRs.
On the other hand, IC losses also compete with synchrotron losses that in turn might pose a challenge to use the FRC as SFR estimator, in particular at high redshifts, where IC losses due to scattering of CR electrons off CMB photons are expected to be larger. Thus, in addition to the question of calorimetry, understanding the relative importance of IC and synchrotron losses is essential.  

To model these processes in more detail, we perform MHD simulations with the moving-mesh code \arepo, simulate the CR proton energy density and model the non-thermal CR spectra. This is the third and last paper of a series in which we present our results. The first paper \citep[][hereafter Paper I]{2021WerhahnI} details the steady-state modeling of CRs in our simulations and compares the results to CR data. We find very good agreement for our CR electron and proton spectra as well as the positron fraction as a function of particle energy when compared to Voyager-1 and AMS-02 data. The resulting gamma-ray emission of star-forming galaxies is analysed in the second paper \citep[][hereafter Paper II]{2021WerhahnII}, where we successfully match the observed FIR-gamma-ray relation as well as observed gamma-ray spectra of star-forming galaxies. Here, we apply the same formalism as described in \citetalias{2021WerhahnI} and study the resulting radio synchrotron emission from primary and secondary CR electrons. In a companion paper, \citet{2021Pfrommer} investigate the origin of the global FRC. In particular, the role of the turbulent small-scale dynamo is identified and the processes leading to the observed scatter in the FRC are analyzed.

This paper is structured in the following way. In Section~\ref{sec: methods}, we describe our simulations of isolated galaxies, the steady-state modeling of the CR proton, primary and secondary electron spectra, as well as the calculation of the resulting radio emission and absorption processes. Section~\ref{sec:timescales} provides an overview of the timescales of the non-radiative and radiative losses of CR protons and electrons. We furthermore describe the modelling of the FRC in Section~\ref{Sec: FIR-Radio} where we analyse the primary and secondary contribution to the total radio luminosity, quantify the calorimetry of CR electrons and assess the processes that have been proposed to `conspire' to maintain an almost linear FRC at high SFRs. Eventually, we scrutinise three possible mechanisms, that could be responsible for the observed hard radio spectra of star-forming galaxies such as NGC~253, M82 and NGC~2146 in Section~\ref{sec:radio spectra}, before we summarise our results in Section~\ref{sec: conclusions}.

\section{Description of the Methods} \label{sec: methods}
\subsection{Simulations}
We perform simulations with the moving mesh code \arepo \citep{2010Springel,2016aPakmor} as described in \citetalias{2021WerhahnI} and \citetalias{2021WerhahnII}. Our simulations start from the gravitational collapse of a gas cloud embedded in a dark matter (DM) halo prescribed by an NFW \citep{1997Navarro} profile, with masses ranging from $M_{200}=10^{10}$ to $10^{12}\,\mathrm{M_{\odot}}$, in order to resemble realistic halo masses from dwarf to Milky Way (MW) sized galaxies. 
Subsequently, we simulate the formation of a rotationally supported disc that forms stars stochastically above a critical density threshold \citep{2003SpringelHernquist}. After an initial burst of star formation, the SFR decreases exponentially over time. At SNe, CRs are instantaneously injected with an energy fraction $\zeta_{\mathrm{SN}}$ of the kinetic energy of the SN explosion, which we vary from 5 to 10 per cent. The lower range of $\zeta_{\mathrm{SN}}$ is inferred from a combination of kinetic plasma simulations at oblique shocks \citep{Caprioli2014} and three-dimensional MHD simulations of CR acceleration at SN remnant shocks \citep{Pais2018}, which is followed by a detailed comparison of simulated radio, X-ray and gamma-ray emission maps and spectra to observational data \citep{Pais2020a,Pais2020b,Winner2020}. 

The ideal MHD prescription \citep{2013Pakmor} governs the evolution of the magnetic field. Starting from an initial seed magnetic field $B_0$ permeating the gas cloud before collapse, the magnetic field is exponentially amplified by a small-scale dynamo, before it grows further via an inverse cascade until saturation \citep{2021Pfrommer}. We chose two different values for the initial magnetic field of $B_0 = 10^{-10}$ and $10^{-12}\,\mathrm{G}$, which represent the pre-amplified magnetic field in
a proto-galactic environment. CRs are described as a relativistic fluid \citep{2016aPakmor, 2017aPfrommer} with adiabatic index of 4/3 and we account for adiabatic changes of the CR energy density as well as Coulomb and hadronic loses due to interactions with the ISM. Furthermore, CRs are advected with the gas (in our `CR adv' models) and are additionally allowed to anisotropically diffuse along magnetic field lines (in our `CR diff' models) with a parallel diffusion coefficient along magnetic field lines of $D_0=10^{28}\mathrm{cm^2\,s^{-1}}$. 

A summary of all simulations analysed in this paper is presented in Table~\ref{Table-Simulations}. The concentration parameter of the NFW profile is given by $c_{200}=r_{200}/r_s=12$ in all simulations, where the characteristic scale radius of the NFW profile is denoted by $r_s$ and $r_{200}$ is the radius enclosing a mean density that is equal to 200 times the critical density of the universe. 
We refer to the simulation in the `CR diff' model with a halo mass of $M_{200}=10^{12}\,\mathrm{M_\odot}$, $B_0=10^{-10}$~G and $\zeta_\mathrm{SN}=0.05$ as `fiducial halo' in the following.

\begin{table}
 \caption{Overview of the simulations in this paper. Shown are (1) the halo mass $M_{200}$, (2) the CR transport model: in the `CR adv' model we only account for CR advection with the gas whereas the `CR diff' model additionally allows for anisotropic diffusion, (3) the initial magnetic field $B_0$, (4) the injection efficiency of CRs at SNRs, $\zeta_\mathrm{SN}$, and (5) the referenced figures.}
 \label{Table-Simulations}
 \begin{threeparttable}[t]
 \begin{tabular}{lcccc}
  \hline
  $M_{200} [\mathrm{M_\odot}]$  & CR model  & $B_0$  $\mathrm{[G]}$ & $\zeta_{\mathrm{SN}}$ & Figures\\
  \hline
  \hline
 $10^{10}$            & CR adv                  &  $10^{-10}$  & 0.05 & \ref{fig: FIR-Radio}, \ref{fig: calorimetric_fraction}, \ref{fig: delta_q}, \ref{fig: ratios_timescsales_sw} \\
 $10^{11}$            & CR adv / CR diff  &  $10^{-10}$  & 0.05 & \ref{fig: FIR-Radio}, \ref{fig: calorimetric_fraction}, \ref{fig: delta_q}, \ref{fig: ratios_timescsales_sw} \\
 $3\times 10^{11}$    & CR adv / CR diff  &  $10^{-10}$  & 0.05 & \ref{fig: FIR-Radio}, \ref{fig: calorimetric_fraction}, \ref{fig: delta_q}, \ref{fig: ratios_timescsales_sw}\\
 $10^{12}$           & CR adv / CR diff  &  $10^{-10}$  & 0.05 & \ref{fig: tau} to \ref{fig: synchr_alpha_maps}\\
    \hline 
  $10^{10}$            & CR adv                 &  $10^{-12}$  & 0.05 & \ref{fig: FIR-Radio-variations} \\
 $10^{11}$            & CR adv / CR diff  &  $10^{-12}$  & 0.05 &\ref{fig: FIR-Radio-variations}  \\
 $3\times 10^{11}$    & CR adv / CR diff  &  $10^{-12}$  & 0.05 & \ref{fig: FIR-Radio-variations}  \\
 $10^{12}$           & CR adv / CR diff  &  $10^{-12}$  & 0.05 &\ref{fig: FIR-Radio-variations}   \\
   \hline
  $10^{10}$            & CR adv                 &  $10^{-12}$  & 0.10 &\ref{fig: FIR-Radio-variations} \\
 $10^{11}$            & CR adv / CR diff  &  $10^{-12}$  & 0.10 &\ref{fig: FIR-Radio-variations} \\
 $3\times 10^{11}$    & CR adv / CR diff  &  $10^{-12}$  & 0.10 & \ref{fig: FIR-Radio-variations} \\
 $10^{12}$           & CR adv / CR diff  &  $10^{-12}$  & 0.10 &\ref{fig: FIR-Radio-variations}   \\
\hline
 \end{tabular}
 \end{threeparttable}
\end{table}

\subsection{CR steady-state spectra}

In addition to the primary CR electron population that is accelerated at SNRs together with CR protons, hadronic interactions of CR protons with the ISM lead to the production of neutral and charged pions. Pions decay further into muons and eventually to neutrinos and secondary electrons and positrons (hereafter referred to as secondary electrons). 

In order to model the spectral distribution of CRs, we assume a steady state and solve the diffusion-loss equation \citep{1964ocr..book.....G,2004Torres} in each cell for protons, primary and secondary electrons (see \citetalias{2021WerhahnI}), which reads
\begin{align}
\frac{\mathrm{}f_i(E_i)}{\tau_{\mathrm{esc}}}-\frac{\mathrm{d}}{\mathrm{d}E_i}\left[f_i(E_i)b(E_i)\right]=q_i(E_i).
\label{eq:diff-loss-equ}
\end{align}
Here, $E_i$ denotes the CR energy, the subscript $i=\mathrm{e,p}$ specifies the CR species and $f_i(E_i)=\mathrm{d}N_i/(\mathrm{d}V\mathrm{d}E_i)$ is the resulting equilibrium spectral density for either protons (denoted by p), primary or secondary electrons (denoted by e). We define the source function $q_i(E_i)=q_i(p_i)\mathrm{d}p_i/\mathrm{d}E_i$ in terms of a power-law in momentum with an exponential cut-off given by 
\begin{align}
q_{i}(p_{i})\mathrm{d}p_{i} = C_{i} p_{i}^{-\alpha_{\mathrm{inj}}} \exp[-(p_i/p_{\mathrm{cut},i})^{n}]\mathrm{d}p_{i},
\label{eq: source fct. Q(p)}
\end{align}
where $p_i=P_i/(m_i c)$ denote normalised momenta, $m_i$ is the proton/electron rest mass and $c$ is the speed of light. We adapt $n=1$ for protons and $n=2$ for primary electrons \citep{2007Zirakashvili,2010Blasi} in Eq.~\eqref{eq: source fct. Q(p)} and assume cutoff momenta for protons $p_{\mathrm{cut,p}}=1\,\mathrm{PeV}/(m_{\mathrm{p}}c^2)$ \citep{Gaisser1990} and for electrons $p_{\mathrm{cut,e}}=20\,\mathrm{TeV}/(m_{\mathrm{e}}c^2)$
\citep{Vink2012}. Both, primary electrons and protons are injected with a spectral index of $\alpha_{\mathrm{inj}} = 2.2$. We discuss the source function of secondary electrons in Section~\ref{sec:primary and secondary CR electrons}. 

The resulting steady-state distribution of CR protons is ensured to match the CR energy density in each cell, after all cooling processes have been taken into account. The energy loss processes are given by the cooling rate $b(E_i)=-\mathrm{d}E_i/\mathrm{d}t$, that comprise hadronic and Coulomb losses in the case of CR protons. Additionally, losses due to particle escape are quantified by the timescale $\tau_{\mathrm{esc}}=(\tau_{\mathrm{adv}}^{-1}+\tau_{\mathrm{diff}}^{-1})^{-1}$, where advection and diffusion losses are included. These are estimated via
\begin{align}
\tau_{\mathrm{diff}}=\frac{L_{\mathrm{CR}}^{2}}{D}
\label{eq:tau_diff}
\end{align}
and
\begin{align}
\tau_{\mathrm{adv}}=\frac{L_{\mathrm{CR}}}{\varv_{\mathrm{z}}}.
\label{eq:tau_adv}
\end{align}
Here, $L_{\mathrm{CR}}=\varepsilon_{\mathrm{CR}}/\left|\nabla\varepsilon_{\mathrm{CR}}\right|$ is an estimate for the diffusion length. For diffusion losses, we assume an energy dependence of the diffusion coefficient $D=D_0 (E/E_0)^{\delta}$, where $D_0=10^{28}\mathrm{cm^2\,s^{-1}}$, $E_0=3\,\mathrm{GeV}$ and $\delta = 0.5$, which can be inferred from observations of beryllium isotope ratios \citep{2020Evoli}. However, we find in \citetalias{2021WerhahnII}, that gamma-ray spectra of highly star-forming galaxies provide a better match for a shallower energy dependence of the diffusion coefficient with $\delta=0.3$ and hence, we also adopt both values of $\delta$ here. For advection losses, only the verical velocity component $\varv_z$ pointing away from the disc is taken into account, because we show in \citetalias{2021WerhahnI}, that azimuthal fluxes in and out of a computational cell nearly compensate each other, so that advection predominantly happens in vertical direction.

The assumption of a cell-based steady state has been analyzed in \citetalias{2021WerhahnI}, where we compare the timescale of the change in the total energy density of CRs over a global simulation timestep $\tau_{\mathrm{CR}}$ to the loss timescale that includes cooling and diffusion losses $\tau_{\mathrm{all}}$. We find that the requirement for a steady-state configuration, i.e.\ $\tau_\mathrm{all}/\tau_\mathrm{CR}<1$ is reached in cells that contribute most to the non-thermal emission processes, i.e.\ both in the radio and gamma-ray regime. Hence, we consider this a good assumption for our study concerning the non-thermal emission processes. However, because the steady-state assumption breaks down in low density regions, in outflows and near SNRs, we aim towards a more accurate treatment of the time evolution of CR protons and electrons \citep{2020MNRAS.491..993G,2019MNRAS.488.2235W} in future work.

\subsection{Primary and secondary CR electrons\label{sec:primary and secondary CR electrons}}

To obtain the steady-state distribution $f_\mathrm{e}$ of primary CR electrons, we adapt Eq.~(\ref{eq: source fct. Q(p)}) as the source function and solve the steady-state equation, i.e., Eq.\,\eqref{eq:diff-loss-equ}. The normalization of $f_\mathrm{e}$ is set by requiring the primary CR electrons to reproduce the observed ratio of electrons to protons, $K_{\mathrm{ep}}^{\mathrm{obs}}$, at 10~GeV in a snapshot of our simulations resembling the MW in terms of halo mass and SFR. To this end, we average over the cell-based steady-state spectra around the solar galacto-centric radius and re-normalise the spectra according to the observed electron to proton ratio. From this, we can infer the injected ratio of primary electrons to protons before cooling, $K_{\mathrm{ep}}^{\mathrm{inj}}\approx 0.02$, that is needed, in order to obtain the observed value of  $K_{\mathrm{ep}}^{\mathrm{obs}}\approx 0.01$ at 10~GeV \citep{Cummings2016} after cooling. We assume that the injected ratio of electrons to protons is universal and adapt it to the rest of the galaxy as well as all other simulated galaxies.

To calculate the production of secondary CR electrons and positrons, we adopt the parametrization of \cite{2006PhRvD..74c4018K} for large kinetic proton energies $T_{\mathrm{p}}>100\,\mathrm{GeV}$. Because \cite{2018Yang} provide a more detailed modeling of the differential cross section of pion production near the threshold of pion production up to $10\,\mathrm{GeV}$, we use their parametrization for small proton energies and perform a cubic spline interpolation in the intermediate energy range, combining it with our own fit of the total cross section of pion production to the data (see equations B1, B5 and B6 in \citetalias{2021WerhahnI}). 

In addition to Coulomb and escape losses, CR electrons also suffer losses due to radiative processes, including synchrotron, bremsstrahlung and IC emission (see Section~\ref{sec:timescales}). In order to model the latter, we assume that the energy density of the incident radiation field is composed of the CMB and stellar radiation, i.e.\ $\varepsilon_{\mathrm{ph}}=\varepsilon_{\mathrm{CMB}}+\varepsilon_{\star}$. The stellar contribution can be approximated by the FIR emission that results from dust-reprocessed UV light from young stellar populations, which we approximate with a black body distribution characterised by a temperature $T=20$~K \citep{2000Calzetti}. The resulting photon energy density in each cell is calculated as 
\begin{equation}
\varepsilon_{\star}=\sum_{i}\frac{L_{\mathrm{FIR},i}}{4 \upi R_{i}^{2}c},
\label{eq:photon-energy-density}
\end{equation}
where we infer the FIR-luminosity $L_{\mathrm{FIR},i}$ from the SFR in each cell \citep{1998Kennicutt} and sum over the fluxes arriving from all other cells at a distance of $R_i$ by using a tree-code. If a cell is itself actively star forming, we equate the distance with the cells' radius, which we derive from its volume $V_i$ via $R_i=[3V_i /(4\pi)]^{1/3}$.

Of particular interest in this work is the relative importance of radio emission due to primary vs. secondary CR electrons. The ratio of their distribution functions at 10~GeV has been derived in \citetalias{2021WerhahnI} using a simplified analytical approximation:
\begin{align}
\frac{f_{\mathrm{e}}^{\mathrm{prim}}}{f_{\mathrm{e}}^{\mathrm{sec}}} &  =K_{\mathrm{ep}}^{\mathrm{inj}}\frac{3}{128}16^{\alpha_{\mathrm{p}}}\left(1+\frac{\tau_{\pi}}{\tau_{\mathrm{esc}}}\right),
 \label{eq:N_e_prim/N_e_sec}
\end{align}
where $\tau_{\pi}$ denotes the timescale of pion production via hadronic CR proton interactions with the ISM, $\alpha_{\mathrm{p}}$ is the spectral index of the CR proton distribution, and $f_{\mathrm{e}}^{\mathrm{sec}} = f_{\mathrm{e}^+}^{\mathrm{sec}} + f_{\mathrm{e}^-}^{\mathrm{sec}}$ is the total steady-state distribution of secondary electrons and positrons. Adapting $K_{\mathrm{ep}}^{\mathrm{inj}}=0.02$ and calorimetric conditions for secondary electron production, i.e. $\tau_{\pi}\ll\tau_{\mathrm{esc}}$ ($\tau_{\pi}\simeq \tau_{\mathrm{esc}}$), which consequently implies $\alpha_{\mathrm{p}}=\alpha_{\mathrm{inj}}=2.2$, Eq.~\eqref{eq:N_e_prim/N_e_sec} yields $f_{\mathrm{e}}^{\mathrm{prim}}/f_{\mathrm{e}}^{\mathrm{sec}}\approx 0.2$ ($0.4$). 
In fact, this can be interpreted as a lower limit, because as soon as escape losses become more important than hadronic losses, i.e.\ $\tau_{\pi}\gtrsim\tau_{\mathrm{esc}}$, secondary production ceases to be efficient and primary electrons will dominate over secondary electrons and positrons. Thus, we expect for simulations that only account for CR advection and neglect CR diffusion secondary electrons to dominate by a factor of $\sim5$ at 10~GeV in the calorimetric limit. By contrast, if we also include energy dependent diffusion, we will obtain steeper CR proton spectra with $\alpha_{\mathrm{p}}\gtrsim 2.2$ as soon as diffusion losses become important. This increases the ratio of primary to secondaries at 10~GeV significantly and we can expect primary electrons to dominate over secondaries.

\subsection{Radio emission and absorption processes}

Radio emission from star-forming galaxies is due to non-thermal synchrotron as well as thermal bremsstrahlung emission, i.e\ free-free emission. The modeling of these emission processes, together with the corresponding absorption processes, is summarised in the following. 

\subsubsection{Radio synchrotron emission\label{sec: radio synchrotron emission}}

Using our steady-state CR electron population $f_\mathrm{e}$, we calculate the radio synchrotron emissivity, $j_{\nu}=E\, \mathrm{d}N_{\gamma}/(\mathrm{d}\nu \mathrm{d}V \mathrm{d}t)$, in each computational cell following \citet{1986rpa..book.....R}, via
\begin{align}
j_{\nu}=\frac{\sqrt{3}e^{3}B_{\perp}}{m_{e}c^{2}}\intop_{0}^{\infty}f_{\mathrm{e}}(p_{\mathrm{e}})F(\nu/\nu_\rmn{c})\mathrm{d}p_{\mathrm{e}},
\label{eq: j_nu synchr.}
\end{align}
where $B_\perp$ is the component of the magnetic field perpendicular to the line of sight, $e$ denotes the elementary charge and we use an analytical approximation provided by \citet{2010PhRvD..82d3002A} for the dimensionless synchrotron kernel $F(\nu/\nu_\rmn{c})$ (Eq.~\ref{eq: F(x)}), where $\nu_\rmn{c}$ is the critical frequency (see Appendix~\ref{App: Synchrotron emission} for details).
The typical synchrotron emission frequency $\nu_\mathrm{syn}\approx 2\nu_\mathrm{c}$ is related to the electron Lorentz factor $\gamma_\mathrm{e}$ via\footnote{This relation can be derived by approximating the synchrotron kernel with a Dirac delta distribution, but see also App.~\ref{App: Synchrotron emission}.}
\begin{equation}
  \label{eq: nu_synchr}
  \nu_\rmn{syn}
  =\frac{3eB}{2\upi\,m_{\rmn{e}}c}\,\gamma_\rmn{e}^2\simeq1~\rmn{GHz}\,\frac{B}{1\,\umu\mbox{G}}\,
  \left(\frac{\gamma_\mathrm{e}}{10^4}\right)^2.
\end{equation}
This indicates that observations of synchrotron emission at a fixed frequency typically probe electrons with lower (higher) Lorentz factors in higher (lower) magnetic field strengths, which we refer to as the `$\nu_\mathrm{c}$-effect' in the following.
The specific radio synchrotron intensity $I_{\nu}$ at frequency $\nu$ (in units of $\mathrm{erg\,cm^{-2}\,s^{-1}\,Hz^{-1}\,sterad^{-1}}$) is obtained by integrating $j_{\nu}$ along the line of sight (denoted by $s$) and the specific radio luminosity (in units of $\mathrm{erg\,s^{-1}\,Hz^{-1}}$) is obtained by integrating $j_{\nu}$ over the entire galaxy, yielding
\begin{align}
I_\nu = \frac{1}{4\upi} \intop_0^{\infty}  j_\nu \mathrm{d}s
\qquad\mbox{and}\qquad
L_{\mathrm{\nu}}=\intop  j_\nu \mathrm{d}V.
\label{eq:I_nu definition}
\end{align}

\subsubsection{Thermal free-free emission and absorption processes \label{sec: free-free-emission and absorption}}

In addition to non-thermal synchrotron emission, we also expect a contribution from thermal free-free emission in the radio band, which predominantly depends on the electron density and temperature. To model this accurately, a multi-phase model of the ISM with radiative transfer would be required. Since this is beyond the scope of this work, we adapt a simplified model here, where we assume a temperature of $T=8000$~K for the warm ionized medium, which dominates this emission component. To estimate the corresponding electron density, we take a fixed fraction $\xi_\rmn{e}$ of the electron density provided by our simplified pressurised ISM model \citep{2003SpringelHernquist}, that models the ISM with an effective equation of state, and leave $\xi_\rmn{e}$ as a free parameter of order unity.

In addition, we account for free-free absorption and synchrotron self-absorption by means of the radiative transfer equation (see Eq.~\ref{eq: I_nu rad. transfer}). To this end, we (i) rotate the simulation into a desired inclination of the disc with the line of sight, (ii) construct thinly-spaced slices through the simulation box perpendicular to the line of sight, from the front of the simulation volume to its back, and (iii) cumulatively add the optical depth, that is a sum of the optical depth of free-free absorption, $\tau_{\mathrm{ff}}$, and synchrotron self-absorption, $\tau_{\mathrm{SSA}}$. Because the electron density along the line of sight depends on galactic inclination, the resulting amount of free-free emission and absorption inherits this dependence and so does the shape of the radio spectrum.
Whereas free-free absorption predominantly affects the low-frequency part of radio spectra (see Section~\ref{sec: radio spectra: thermal and non-thermal radio emission}), we find that synchrotron self-absorption has a negligible effect at radio frequencies studied here.

\section{Timescales} \label{sec:timescales}
This section will provide an overview of the timescales of all relevant processes, which we define as $\tau=E/\dot{E}$. Because we aim at explaining the FRC at 1.4~GHz, we compute the timescales at a fixed observational frequency. According to Eq.~(\ref{eq: nu_synchr}), electrons that emit synchrotron radiation at GHz frequencies in $\umu$G magnetic fields have a typical energy of $E_{\mathrm{e}}=10^4\,m_{\mathrm{e}}c^2\approx 5$~GeV. These electrons can be primary and secondary electrons while the latter have been hadronically created by CR protons with typical energies of $E_{\mathrm{p}}\approx 16 E_\mathrm{e}\approx 80~$GeV.

\subsection{Non-radiative processes}

\begin{figure*}
\begin{centering}
\includegraphics[scale=1]{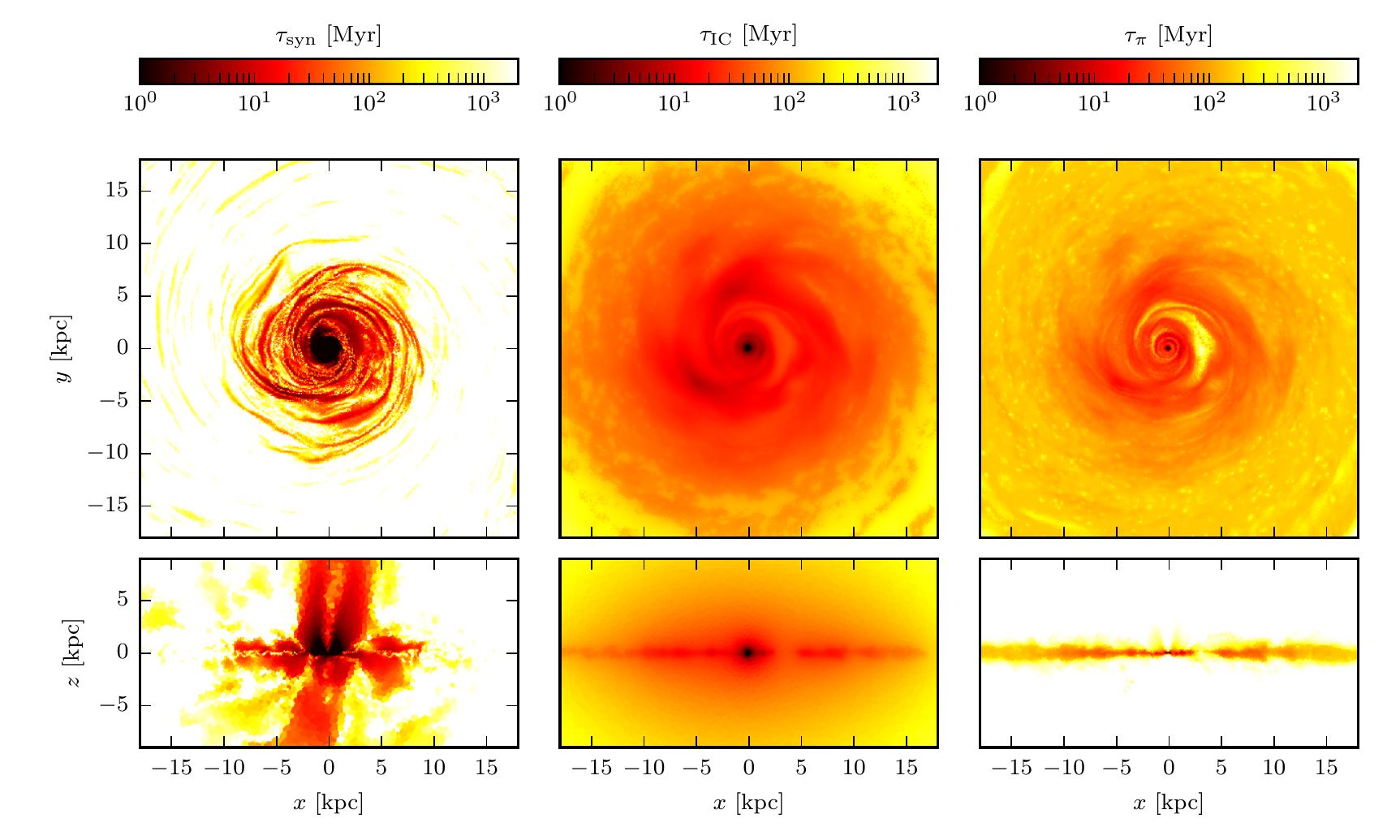}
\includegraphics[scale=1]{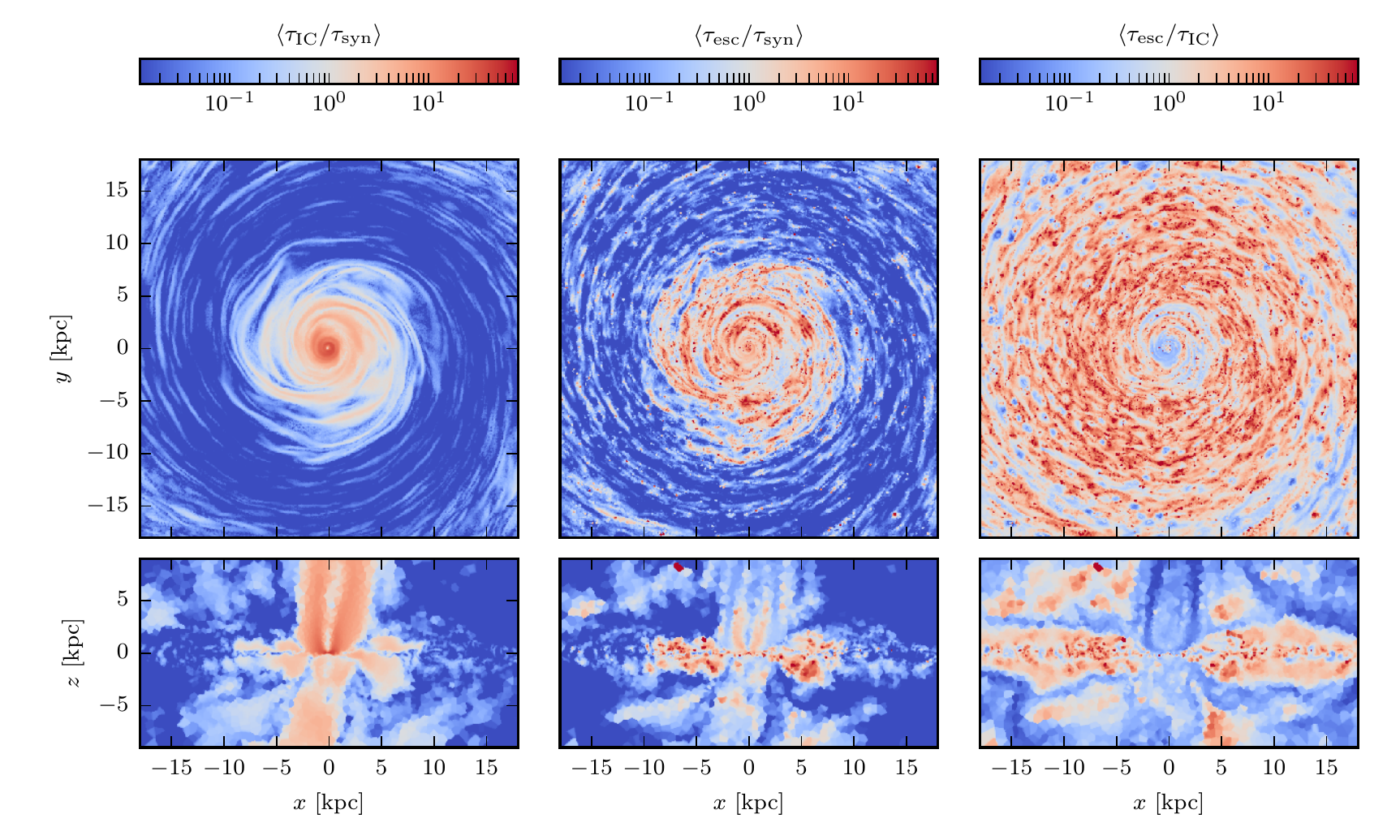}
\par\end{centering}
\caption{The upper panels show from left to right maps (slices) of the characteristic cooling timescales of synchrotron and IC cooling for CR electrons, as well as the timescale of hadronic interactions of CR protons, that is relevant for the production of secondary electrons. In the lower panels, we show the ratios of IC to synchrotron cooling, as well as the ratio of escape to synchrotron and to IC losses, respectively, where escape losses include losses due to CR advection and diffusion. The timescale ratios are averaged over thin slices with a thickness of 500~pc and the timescales are calculated at an energy of 10~GeV. All maps are shown for a snapshot with a halo mass of $M_{200}=10^{12}\,\mathrm{M_{\odot}}$ at $t=2.3$~Gyr.}
\label{fig: tau}
\end{figure*}

The injection timescale of secondaries corresponds to the timescale of hadronic interactions of CR protons with the ISM, and is given by
\begin{align}
    \tau_{\pi} = 
    \frac{1}{c n_{\mathrm{N}}K_{\mathrm{p}}\sigma_{\mathrm{pp}}}
    \approx 240 \, \left(\frac{n_{\mathrm{N}}}{0.1~\rmn{cm}^{-3}}\right)^{-1}\mathrm{Myr},
    \label{eq: tau_pi}
\end{align}
where the cross section of hadronic interactions is $\sigma_{\mathrm{pp}}\approx 44 \,\mathrm{mbarn}$ for $\alpha_{\mathrm{p}}=2.2$ \citep{2017bPfrommer}, the inelasticity of hadronic interactions is given by $K_{\mathrm{p}}=1/2$ \citep{1994A&A...286..983M}, $n_{\mathrm{N}}=n_{\mathrm{H}} + 4n_{\mathrm{He}} =(X_{\mathrm{H}} + 1- X_{\mathrm{H}})\rho/m_{\mathrm{p}}=\rho/m_{\mathrm{p}}$ is the number density of target nucleons in the ISM, where $X_{\mathrm{H}}=0.76$ denotes the hydrogen fraction and $\rho$ is the gas density. 
Consequently, the hadronic timescale directly traces the gas density, as can bee seen in the upper right panels in Fig.~\ref{fig: tau}, which shows $\tau_{\pi}$ for our M82-like galaxy, see Table \ref{Table-Galaxies}.

Coulomb interactions of CRs with the ambient medium of an electron number density $n_\mathrm{e}$ act on a timescale \citep{GOULD1972145}
\begin{align}
    \tau_\mathrm{Coul}  &= \frac{2E_\mathrm{e} \beta_{\mathrm{e}}}{{3\sigma_\rmn{T}n_{\mathrm{e}}m_{\mathrm{e}}c^3}} \left[\ln\left(\frac{m_{\mathrm{e}}c^{2}\beta_{\mathrm{e}}\sqrt{\gamma_{\mathrm{e}}-1}}{\hbar\omega_{\mathrm{pl}}}\right)-\right.\nonumber \\
 & \left.\ln\left(2\right)\left(\frac{\beta_{\mathrm{e}}^{2}}{2}+\frac{1}{\gamma_{\mathrm{e}}}\right)+\frac{1}{2}+\left(\frac{\gamma_{\mathrm{e}}-1}{4\gamma_{\mathrm{e}}}\right)^{2}\right]^{-1} 
    \propto \frac{E_\mathrm{e}}{n_\mathrm{e}}\propto \frac{1}{B^{1/2} n_\mathrm{e}},
    \label{eq: tau_coul}
\end{align}
where the normalised electron velocity is denoted by $\beta_\mathrm{e}=\varv_\mathrm{e}/c $, $\sigma_{\mathrm{T}}$ is the Thompson cross section, the plasma frequency is defined by $\omega_\mathrm{pl}=\sqrt{4\upi e^{2}n_{\mathrm{e}}/m_{\mathrm{e}}}$ and we used in the last step the relativistic limit and Eq.~(\ref{eq: nu_synchr}) so that the last expression is only valid at fixed synchrotron emission frequency. Adapting an electron number density of $n_\mathrm{e}=0.1\,\mathrm{cm^{-3}}$, Coulomb losses act on timescales of $\sim 2~\mathrm{Gyr}$ for highly relativistic electrons with $\gamma_\mathrm{e}=5~\mathrm{GeV}/(m_\mathrm{e}c^2)\approx 10^4$, whereas middly-relativistic electrons with $\gamma_\mathrm{e}\sim 10 $ cool on timescales of $\tau\approx 3\,\mathrm{Myr}$. Thus, Coulomb losses typically remove the low-energy part of the CR electron spectrum. Consequently, in order for radio-synchrotron emitting electrons (with typical Lorentz factors $\gamma_\mathrm{e}\sim 10^{4}$ for $B=1~\umu$G and $\nu_\rmn{syn}=1.4$~GHz) to be affected by Coulomb losses, very high densities are required.

\subsection{Radiative processes}

At high electron energies, radiative losses due to synchrotron, IC or bremsstrahlung emission are typically dominant.
The loss timescale due to synchrotron emission is given by
\begin{align}
    \tau_\mathrm{syn} &=\frac{3E_\mathrm{e}}{\sigma_{\mathrm{T}} c \beta_{\mathrm{e}}^{2} \gamma_{\mathrm{e}}^{2}\varepsilon_{B} }
    \propto \frac{1}{E_\mathrm{e} B^2}\propto B^{-3/2},
    \label{eq: tau_syn}
\end{align}
where $\varepsilon_B=B^2/(8\upi)$ and we used Eq.~\eqref{eq: nu_synchr} in the last step so that this dependence is only valid at fixed synchrotron emission frequency. For instance, in central regions of starburst galaxies with magnetic field strengths of $\approx 10\, \umu$G, where electrons with typical energies of $E_{\mathrm{e}}\approx 2~$GeV are responsible for radio synchrotron emission at GHz frequencies, the synchrotron cooling timescale is $\tau_{\mathrm{syn}}\approx 70\,\mathrm{Myr}$. In starburst nuclei with up to $B\approx 50~\umu$G, synchrotron cooling acts on even shorter timescales of $\tau_\mathrm{syn}\approx 6\,\mathrm{Myr}$.

In an ambient radiation field with a photon energy density $\varepsilon_{\mathrm{ph}}$, CR electrons scatter off of these photons and lose energy on a timescale
\begin{align}
    \tau_\mathrm{IC} &=\frac{3E_\mathrm{e}}{\sigma_{\mathrm{T}} c \beta_{\mathrm{e}}^{2} \gamma_{\mathrm{e}}^{2}\varepsilon_{\mathrm{ph}} }
    \propto \frac{1}{E_\mathrm{e} \varepsilon_\mathrm{ph}}\propto \frac{B^{1/2}}{\varepsilon_\mathrm{ph}},
    \label{eq: tau_IC}
\end{align}
where the last step is only valid at fixed synchrotron emission frequency. Note that $\varepsilon_{\mathrm{ph}}$ is a sum of a stellar contribution ($\varepsilon_{\star}$, modeled by Eq.~\ref{eq:photon-energy-density}) and the CMB ($\varepsilon_{\mathrm{CMB}}\approx 4.16\times 10^{-13}\,(1+z)^2\,\mathrm{erg\,cm^{-3}}$, where $z$ is the cosmic redshift). The CMB is ubiquitous and independent of the local SFR (or $\varepsilon_\star$).
For a photon energy density of $\varepsilon_{\mathrm{ph}}=5\varepsilon_{\mathrm{CMB}}$, the IC cooling timescale ranges from $\sim 40$~Myrs to $\sim 140$~Myrs in regions with magnetic field strengths of $1$ to $10\,\umu$G, where we again adapted Eq.~(\ref{eq: nu_synchr}). 
Because $\tau_\mathrm{syn}/\tau_{\mathrm{IC}}\propto B_\mathrm{ph}^2/B^2$, where $B_{\mathrm{ph}}=(8\upi \varepsilon_{\mathrm{ph}})^{1/2}$ is the equivalent magnetic field strength, IC losses will always dominate over synchrotron losses if $B_{\mathrm{ph}}>B$. Because the photon energy density of the CMB corresponds to $B_{\mathrm{CMB}}\approx 3~\umu\mathrm{G}$ today, IC losses will be relevant as soon as the magnetic field drops below $3~\umu\mathrm{G}$, and correspondingly at larger magnetic field strengths if we additionally account for a stellar radiation field.

In a fully ionized medium with a proton number density $n_{\mathrm{p}}$, CR electrons lose energy due to the emission of bremsstrahlung on a timescale \citep{1970BlumenthalGould}
\begin{align}
    \tau_\mathrm{brems} &= \frac{E_\mathrm{e}}{4\alpha r_{0}^{2}cn_{\mathrm{p}}\beta_{\mathrm{e}}\gamma_{\mathrm{e}}m_\mathrm{e}c^2}
    \left[\ln(2\gamma_{\mathrm{e}})-\frac{1}{3}\right] ^{-1}
    \propto \frac{1}{n_{\mathrm{p}} \ln E_\mathrm{e}},
    \label{eq: tau_brems}
\end{align}
where $\alpha$ is the fine structure constant and $r_0$ denotes the electron radius.
Again, adapting electrons with energies of 5~GeV, bremsstrahlung losses act in a medium with $n_{\mathrm{p}}=0.1\,\mathrm{cm^{-3}}$ on timescales of $\tau_\mathrm{brems} \approx 480~\mathrm{Myr}$. Consequently, high densities are required in order for bremsstrahlung losses to be able to compete with synchrotron or IC losses, as we will further discuss in Section~\ref{sec: radio spectra: bremsstrahlung}.

\subsection{Large dynamic range}

\begin{figure*}
\begin{centering}
\includegraphics[scale=1]{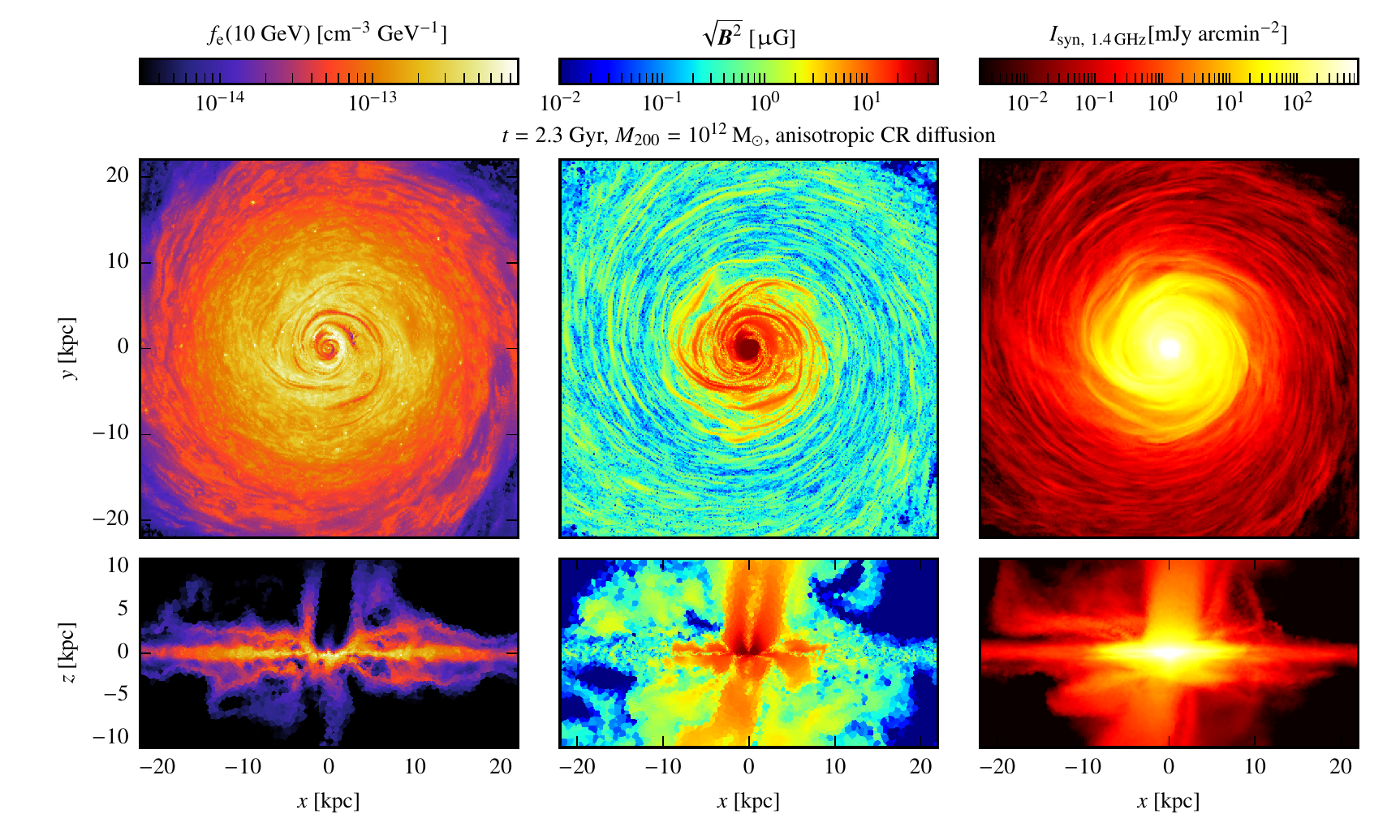}
\par\end{centering}
\caption{Face-on (upper panels) and edge-on (lower panels) maps of the CR electron distribution $f_{\mathrm{e}}$ at 10~GeV (left hand panels), the magnetic field strength (middle panels) and the total (primary and secondary) synchrotron intensity at 1.4~GHz (right hand panels) of a simulation with $M_{200}=10^{12}\,\mathrm{M_{\odot}}$, $B_0=10^{-10}\,$~G and $\zeta_{\mathrm{SN}}=0.05$ at $t=2.3\,$~Gyr  (i.e.\ the same snapshot as shown in Fig.~\ref{fig: tau}).}
\label{fig: maps-properties}
\end{figure*}

The SFR-gas surface density relation, $\dot{\Sigma}_\star\propto\Sigma_\rmn{gas}^{1.4}$, of star-forming and starburst galaxies suggested by \cite{1998Kennicutt} is valid over a large dynamic range of five orders of magnitude in gas surface density. As a result, the CR cooling timescales that depend on gas density are expected to vary on a similarly large range of scales. Because the photon energy density is related to star formation, IC cooling strongly depends on the gas density via the \cite{1998Kennicutt} relation, too. Similarly, the magnetic field strengths of star-forming galaxies are found to scale with gas surface density \citep{2008Robishaw}, resulting in a strong variation of the synchrotron cooling timescale with gas density as well. However, these cooling timescales are not only expected to significantly vary among different types of galaxies, i.e.\ from dwarfs to starbursts, but also within a galaxy. In particular, the relative importance of the cooling and loss timescales is of relevance for shaping their spectra and determining their non-thermal emission properties.

This is exemplified in Fig.~\ref{fig: tau}, where we show maps of the timescale of synchrotron and IC cooling of CR electrons and the hadronic timescale of CR protons, both at an energy of 10~GeV (upper panels), for our fiducial halo at $t=2.3\,\mathrm{Gyr}$. As expected, the synchrotron cooling timescale traces the magnetic field (see Fig.~\ref{fig: maps-properties}) and varies from a few Myr in the very central regions with strong magnetic fields up to a few tens of Myr in the disc at around 5~kpc from the center. At larger radii, the synchrotron timescale increases with decreasing magnetic field as $\tau_\mathrm{syn}\propto B^{-2}$. By contrast, the IC cooling scales as $\tau_\mathrm{IC}\propto \varepsilon_\mathrm{ph}^{-1}$ and consequently, IC cooling remains important at larger radii, where the photon energy density is still high and approaches $\varepsilon_\mathrm{CMB}$. Consequently, IC dominates over synchrotron cooling beyond 5~kpc (see lower left panel in Fig.~\ref{fig: tau}). Note that this ratio is independent of the electron energy, due to the identical energy dependence of IC and synchrotron losses. In those regions where synchrotron cooling dominates over IC cooling, the former is also faster than advection and diffusion losses, i.e.\ escape losses (lower middle panels in Fig.~\ref{fig: tau}). However, in the outskirts of the galactic disc, IC losses dominate over escape losses and only within the outflows, advection losses become relevant.

\section{The FIR-radio correlation}\label{Sec: FIR-Radio}

We will now describe our modelling of the FRC from our simulations, which allows us to dissect the contribution from primary and secondary electrons to the total radio synchrotron luminosity. Furthermore, we will assess calorimetry and analyse possible deviations from the calorimetric picture in starburst galaxies.

\subsection{Modelling the FIR-radio correlation}

\begin{figure*}
\begin{centering}
\includegraphics[scale=1]{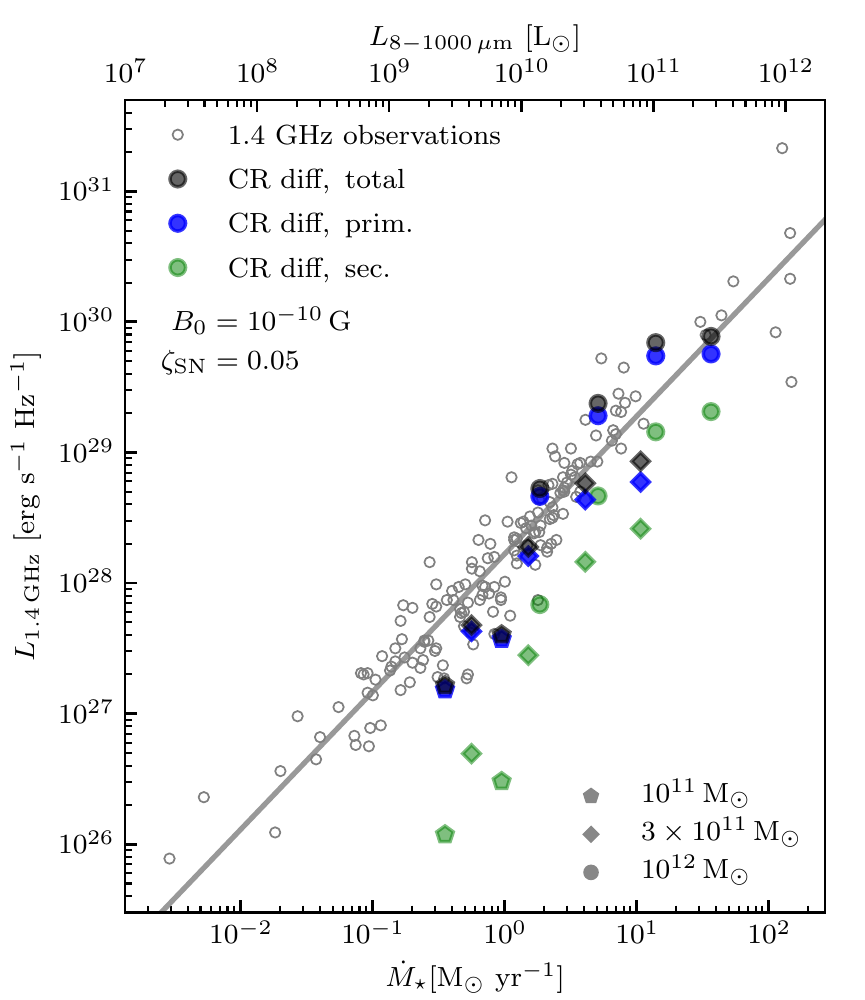}\includegraphics[scale=1]{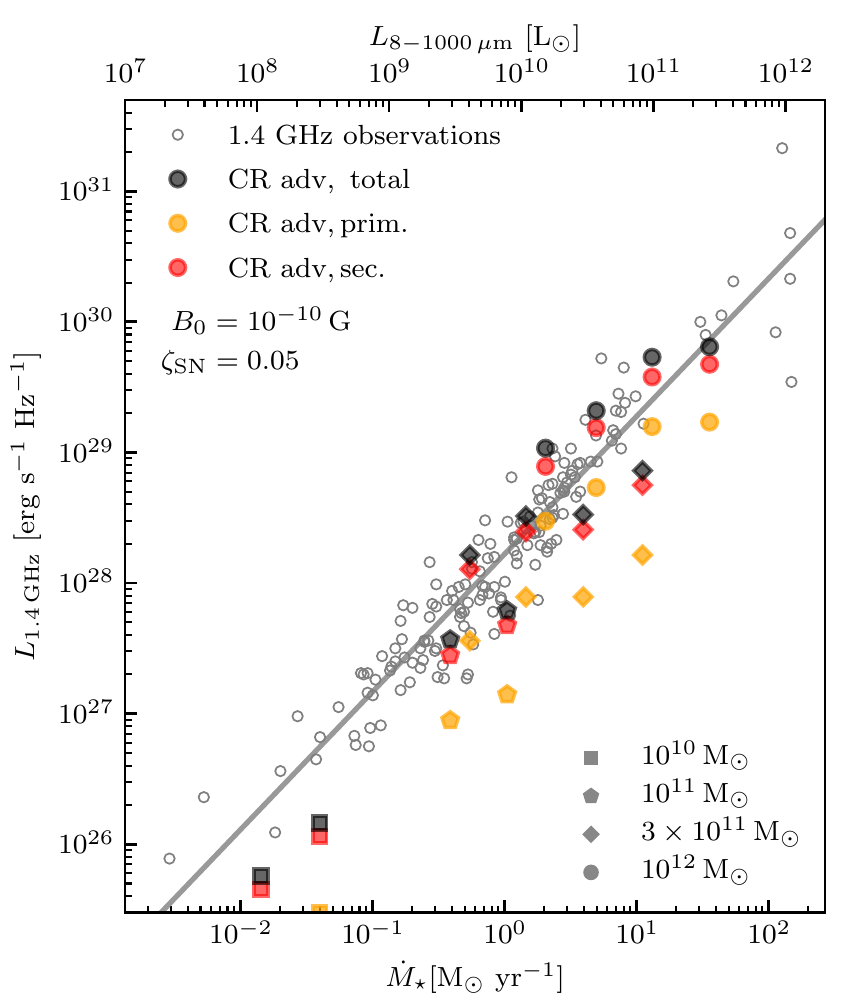}
\includegraphics[scale=1]{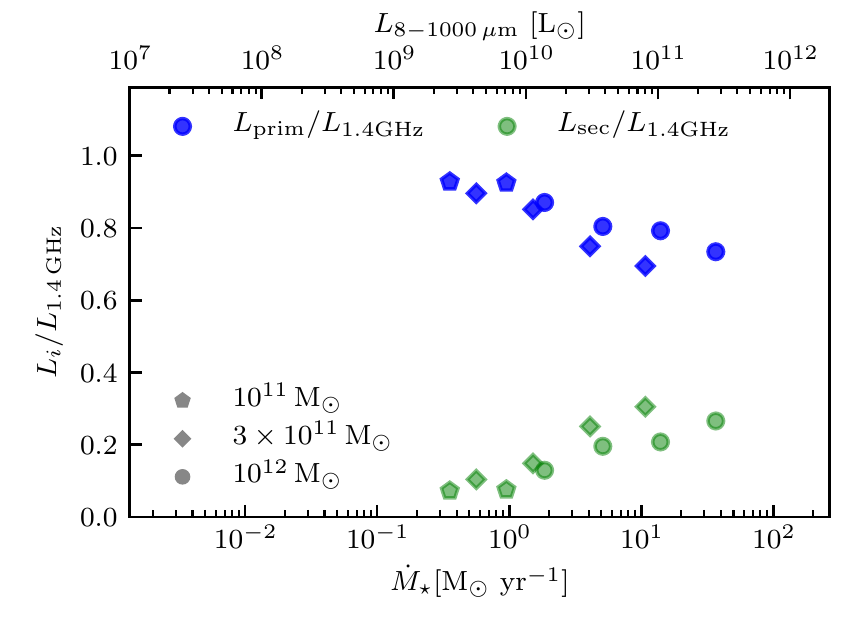}\includegraphics[scale=1]{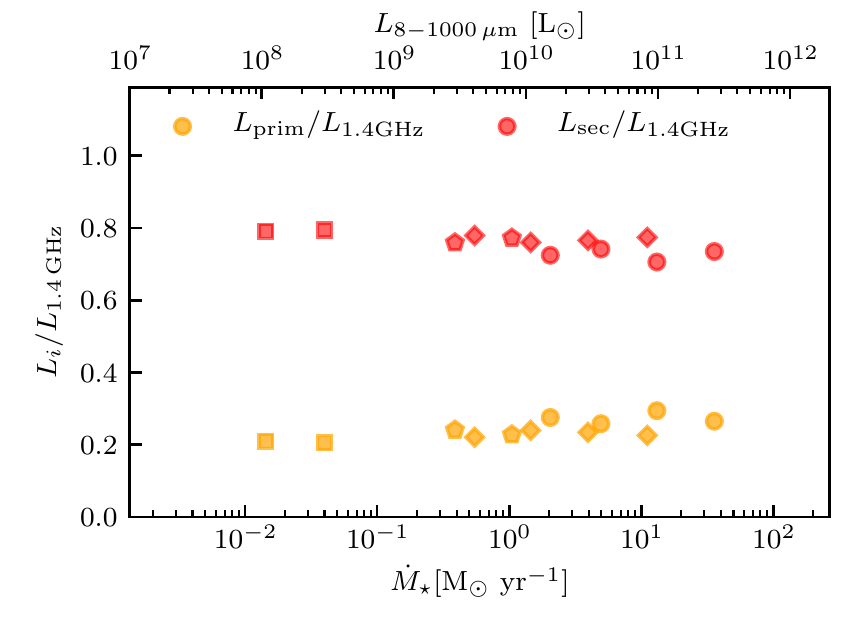}
\par\end{centering}
\caption{Top panels: we compare the observed FRC \citep[][open circles and a fit to the data, grey line]{2003Bell} to our simulated FRC (black symbols for the total radio synchrotron luminosity) in our model that accounts for CR advection and diffusion (`CR diff', left-hand panels) as well as our CR advection-only model (`CR adv', right-hand panels) and assume face-on configurations for all galaxies. Colour coded are contributions from primary (blue and orange) and secondary (green and red) CR electrons and different symbols correspond to simulations with different halo masses (indicated in the legends). All simulations assume $B_0 = 10^{-10}\,\mathrm{G}$ and $\zeta_{\mathrm{SN}}=0.05$. Bottom panels: we show the corresponding relative contributions to the total radio luminosity due to primary and secondary electrons, respectively. Note that the secondary emission dominates in our `CR adv' model while the primary emission dominates in our `CR diff' model. }
\label{fig: FIR-Radio}
\end{figure*}

Post-processing our simulations, we derive the steady-state CR electron distributions $f_\mathrm{e}$ (Eq.~\ref{eq:diff-loss-equ}) in every resolution element. We show the sum of the primary and secondary electron equilibrium distribution at 10~GeV in the left-hand panels of Fig.~\ref{fig: maps-properties} for our fiducial halo, i.e.\ a simulation with halo mass $M_{200}=10^{12}\,\mathrm{M_\odot}$, after 2.3 Gyr. Together with the magnetic field of our MHD simulation (shown in the middle panels of Fig.~\ref{fig: maps-properties}), we can directly calculate the resulting radio synchrotron emission of the steady-state electron distribution, using Eq.~(\ref{eq: j_nu synchr.}). This is shown in the right-hand panels of Fig.~\ref{fig: maps-properties} in terms of the specific radio synchrotron intensity $I_{\nu}$ at a frequency of $\nu=1.4$~GHz as it would be observed in a face-on (upper panel) or edge-on (lower panel) configuration. The radio emission in our projected maps is clearly dominated by the central region up to radii of $\sim$10 kpc, where the magnetic field is strong. Particular filamentary features are visible in the edge-on view of the radio emission. Those trace the morphology of the magnetic field, which reaches up more than 10~kpc above and below the disc, and which is shaped by a strong central outflow driven by the CR pressure gradient.

Because the radio synchrotron emissivity $j_\nu$ in Eq.~(\ref{eq: j_nu synchr.}) depends on the component of the magnetic field perpendicular to the line of sight and hence on the viewing angle, the calculation of the observed radio luminosity also depends on the observed orientation. For the FRC, we chose to calculate the radio luminosity for a face-on configuration of our galaxies. The corresponding edge-on luminosities are typically a factor of about two smaller, which introduces a natural scatter in the FRC \citep{2021Pfrommer}.
We correlate the face-on radio synchrotron luminosities against the SFR ($\dot{M}_{\star}$) of our simulated galaxies with an initial magnetic field of $B_0=10^{-10}$~G and $\zeta_\mathrm{SN}=0.05$ in Fig.~\ref{fig: FIR-Radio}. The corresponding FIR luminosities as derived from the SFRs using the \citet{1998Kennicutt} relation are shown at the upper horizontal axis. For each simulated galaxy, we take snapshots at the peak of the SFR and at times when the SFR has successively decreased by an e-folding, which yields snapshots that are equally spaced in $\log\dot{M}_{\star}$. The effect of varying the initial magnetic field $B_0$ and the injection efficiency of CRs at SNRs, $\zeta_\mathrm{SN}$, on the FRC is discussed in App.~\ref{app: model variation FRC}. Note that we only show the radio synchrotron luminosities here and discuss possible contributions from thermal free-free emission in Section~\ref{sec: radio spectra: thermal and non-thermal radio emission}.

Our simulation models which only allow for CR advection (`CR adv', right-hand panel of Fig.~\ref{fig: FIR-Radio}) and where we additionally account for anisotropic CR diffusion (`CR diff', left-hand panel) match the observed FRC \citep{2003Bell} over a broad range of SFRs. However, the radio luminosities of the smallest halos with $M_{200}=10^{10}\,\mathrm{M_\odot}$ fall short of the FRC in our `CR diff' model and are not visible within the range of radio luminosities shown in Fig.~\ref{fig: FIR-Radio}. As discussed in \citet{2021Pfrommer}, these dwarf galaxies show a slower dynamo growth in comparison to the more massive halos. Additionally, they generate strong outflows that are launched due to the inclusion of anisotropic diffusion of CRs, enabling CRs to escape from the galactic disc, which further reduces the resulting radio emission. By contrast, in the `CR adv' model, the small-scale dynamo efficiently amplifies the magnetic field and CRs accumulate in the galaxy because they cannot diffuse out of the disc by construction. Hence, our dwarf galaxies with $M_{200}=10^{10}\,\mathrm{M_\odot}$ are able to reach the FRC in the `CR adv' model (see Fig.~\ref{fig: FIR-Radio}, right-hand panel). 

In addition to the total radio luminosity, Fig.~\ref{fig: FIR-Radio} also shows the contributions arising from primary and secondary electrons separately, which are further analysed in the following section.

\subsection{Secondary vs.\ primary synchrotron emission \label{sec:secondary vs. primary synchr. emission}}

Complementary to \cite{2021Pfrommer}, here we aim to understand the individual contributions of primary and secondary electrons to the total radio synchrotron luminosity, which enables us to dissect the reasons for our successful reproduction of the observed FRC. First of all, in our approach, the injected CR proton luminosity is given by
\begin{align}
L_{\mathrm{p}}=\zeta_{\mathrm{SN}} \dot{M}_{\star} \epsilon_{\mathrm{SN}} \ ,
\label{eq: L_p1}
\end{align}
where $\epsilon_{\mathrm{SN}}=E_{\mathrm{SN}}/M_{\star} = 10^{51}\,\mathrm{erg}/ (100~\mathrm{M_\odot})=10^{49}\,\mathrm{erg\,M_{\odot}^{-1}}$ is the SN energy released per unit mass \citep[see][]{2017aPfrommer}, where we assume a \citet{2003Chabrier} initial mass function. Adopting our fiducial injection efficiency of CR energy at SN remnants, $\zeta_{\mathrm{SN}}=0.05$ \citep{Pais2018}, and using the \cite{1998Kennicutt} relation between SFR and FIR luminosity, we obtain
\begin{align}
L_{\mathrm{p}}\approx 5.5\times 10^{-4} \, \left(\frac{\zeta_{\mathrm{SN}}}{0.05}\right)\, L_{\mathrm{FIR}}.
\label{eq: L_p2}
\end{align}
The primary electron population obtains a fraction $\zeta_{\mathrm{prim}}$ of the total luminosity of CR protons in our modeling that is given by \citepalias[see Appendix~A in][]{2021WerhahnI}
\begin{align}
L_{\mathrm{prim-e}}=\zeta_{\mathrm{prim}}\,L_{\mathrm{p}}=\left( \frac{m_{\mathrm{p}}}{m_{\mathrm{e}}} \right)^{\alpha_{\mathrm{p}} -2}K_{\mathrm{ep}}^{\mathrm{inj}}\,L_{\mathrm{p}}.
\label{eq: L_e_prim,e}
\end{align}
Using $K_{\mathrm{ep}}^{\mathrm{inj}}=0.02$ and $\alpha_\mathrm{p}=2.2$, this yields $\zeta_{\mathrm{prim}}\approx 0.09$.

\begin{figure*}
\begin{centering}
\includegraphics[scale=1]{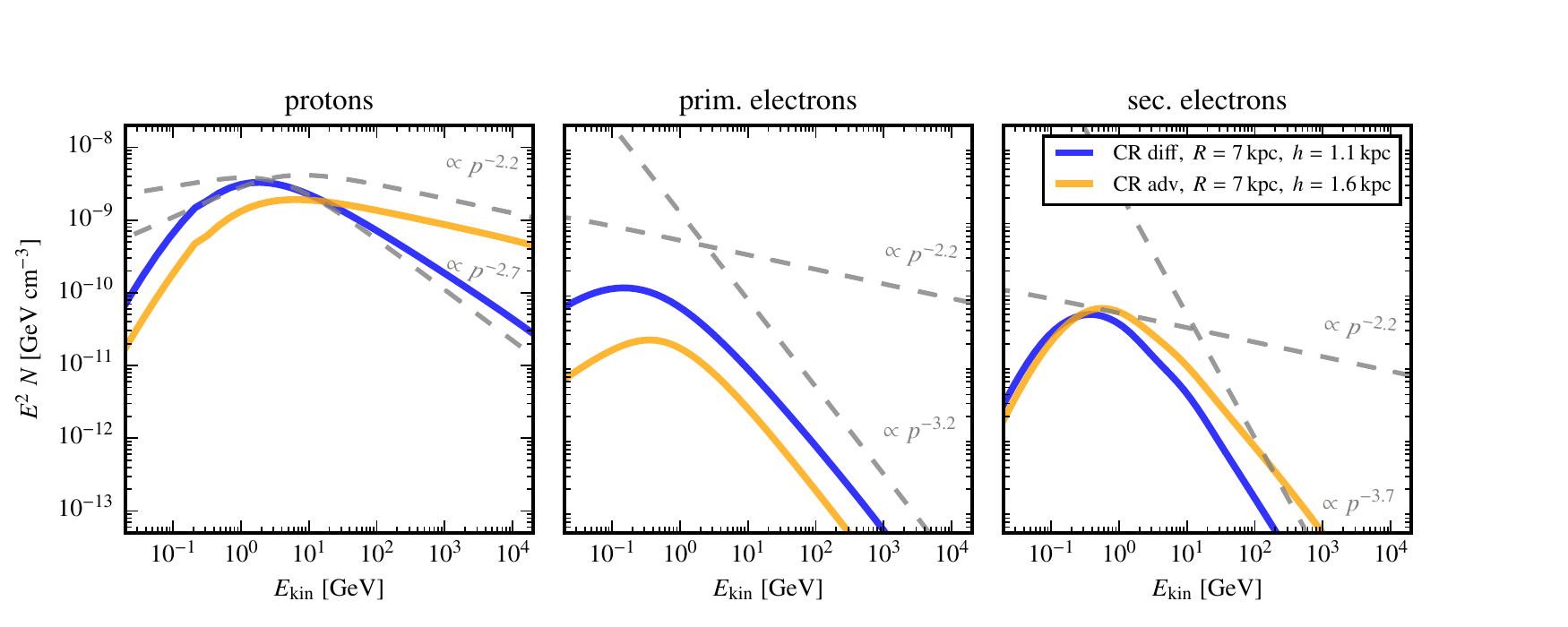}
\par\end{centering}
\caption{From left to right, we show CR proton, primary and secondary electron spectra of simulations with $M_{200}=10^{12}\,\mathrm{M_{\odot}}$, $\zeta_{\mathrm{SN}}=0.05$, $B_0=10^{-10}\,\mathrm{G}$. We contrast the CR spectra of snapshots at $t=1.3\,\mathrm{Gyr}$ of a simulation that includes advection and anisotropic diffusion of CRs (`CR diff', blue) to a simulation that only accounts for CR advection (`CR adv', yellow). In both cases, we average the spectra over the radial range, that includes 99 per cent of the total radio luminosity at 1.4\,GHz (i.e.\ $R\approx 7\,\mathrm{kpc}$ for both snapshots) and the height, where the magnetic field has decreased by an e-folding, as indicated in the legend. The grey dashed lines show momentum power-law spectra (with arbitrary normalisation to exemplify scaling properties) that are typical for the various transport and cooling processes (with indices indicated in the panels). These spectra account for the full proton/electron dispersion relation, which causes the mild downturn of the proton spectrum at energies $E_\rmn{kin}\lesssim 1$~GeV. }
\label{fig: CR-spectra}
\end{figure*}

On the other hand, hadronic interactions of CR protons with the ISM lead to the production of secondary electrons and positrons. Assuming that CR protons exclusively cool via hadronic interactions, the fraction of total proton luminosity injected into secondary electrons and positrons, i.e.\ $\zeta_{\mathrm{sec}}=L_\mathrm{sec-e}/L_{\mathrm{p}}$ can be estimated from the following consideration. Charged pions are produced in hadronic CR proton-proton interactions with a multiplicity $2/3$ and on average, electrons/positrons obtain $1/4$ of the pion energy, which yields a secondary fraction of about $2/3 \times 1/4 =1/6$ of the total proton luminosity. Considering additionally the nuclear enhancement factor of 1.4 to 1.6, which accounts for heavier nuclei in the composition of CRs and the ISM \citep{1976Biallas,1981StephensBadhwar}, we arrive at an energy fraction of $\zeta_{\mathrm{sec}} \approx 0.25$. Hence, secondary electrons and positrons obtain a luminosity of
\begin{align}
L_{\mathrm{sec-e}}&=\eta_{\mathrm{cal,p}}\,\zeta_{\mathrm{sec}}\,L_{\mathrm{p}},
\label{eq: L_e_sec,e}
\end{align}
where we additionally accounted for the calorimetric fraction of protons $\eta_{\mathrm{cal,p}}$. This factor quantifies the fraction of CR proton luminosity that is used for pion production, which reaches values up to $\eta_\mathrm{cal,p}\approx 0.8$ for highly star-forming galaxies (see \citetalias{2021WerhahnII}). We assume a simple power-law momentum spectrum for the source function of CR electrons with a low-momentum cutoff $p_\mathrm{min}$ and calculate the volume-integrated injected electron source function via:
\begin{align}
  \label{eq:Q_e}
  Q_\mathrm{e} (p_\mathrm{e})= \frac{\mathrm{d} N_\mathrm{e}}{\mathrm{d} p_\mathrm{e}\mathrm{d} t}
  = \int q_\mathrm{e} (p_\mathrm{e})\mathrm{d} V
  = \mathcal{C}_\mathrm{e} p_\mathrm{e}^{-\alpha_{\mathrm{e,inj}}}\theta(p_\mathrm{e}-p_\mathrm{min}),
\end{align}
where $\theta (p)$ denotes the Heaviside step function and $\mathcal{C}_{\mathrm{e}}$ is the normalization. Assuming furthermore $\alpha_\mathrm{e,inj}>2$, this enables us to define the total injected CR electron luminosity via
\begin{align}
\label{eq:L_e1}
L_\mathrm{e}&=\int_0^{\infty}Q_\mathrm{e}(p_\mathrm{e})\,T_\mathrm{e}(p_\mathrm{e})dp_\mathrm{e}=\frac{\mathcal{C}_\mathrm{e}\,m_{\mathrm{e}}c^2}{\alpha_\mathrm{e,inj}-1}\\
&\quad\times\left[\frac{1}{2}\,\mathcal{B}_\frac{1}{1+p_\mathrm{min}^2}\left(\frac{\alpha_\mathrm{e,inj}-2}{2},\frac{3-\alpha_\mathrm{e,inj}}{2}\right)+
  p_\mathrm{min}^{1-\alpha_\mathrm{e,inj}}\,
  \frac{T_\mathrm{e}(p_\mathrm{min})}{m_{\mathrm{e}}c^2}\right],\nonumber\\
&\equiv \mathcal{C}_\mathrm{e}\,m_{\mathrm{e}}c^2 A_\mathrm{bol} (p_\mathrm{min},\alpha_\mathrm{e,inj}),
\label{eq:L_e2}
\end{align}
where $T_\mathrm{e}(p_\mathrm{e})=\left(\sqrt{1+p_\mathrm{e}^2}-1\right)\,m_{\mathrm{e}}c^2$ is the kinetic electron energy and $\mathcal{B}_y(a,b)$ denotes the incomplete beta function \citep{1965hmfw.book.....A}. Assuming that the synchrotron cooling time of CR electrons is shorter than their escape time \citep{Voelk1989}, the emitted synchrotron luminosity from a steady-state electron distribution $f_\mathrm{e}$ can be calculated as
\begin{align}
  \label{eq:FRC_L_nu1}
  \nu L_{\nu} (\mathrm{GHz}) &= \frac{E_\gamma\mathrm{d} N_\gamma}{\mathrm{d}\ln\nu\,\mathrm{d} t} 
  =\eta_{\mathrm{syn}}\,\frac{E_\mathrm{e}\mathrm{d}
    N_\mathrm{e}}{2\,\mathrm{d}\ln\gamma_\mathrm{e}\,\mathrm{d} t}
    \approx\eta_{\mathrm{syn}} \, \frac{\gamma_\mathrm{e}^{2-\alpha_\mathrm{e,inj}}}{2A_\mathrm{bol}}\, L_\mathrm{e} 
  \\
  & \equiv \eta_{\mathrm{syn}} \, \zeta_\mathrm{bol}(p_\mathrm{min},\alpha_\mathrm{e,inj},\gamma_\mathrm{e}) \,L_\mathrm{e}.
   \label{eq:FRC_L_nu2}
\end{align}
Here, we use $\mathrm{d}\ln \nu = 2\mathrm{d}\ln \gamma_\mathrm{e}$ (see Eq.~\ref{eq: nu_synchr}) and Eq.~(\ref{eq:L_e2}), while assuming the relativistic limit $\mathrm{d}p_\mathrm{e}\approx\mathrm{d}\gamma_\mathrm{e}$ in Eq.~(\ref{eq:Q_e}). Furthermore, we define the bolometric electron fraction $\zeta_\mathrm{bol} = \gamma_\mathrm{e}^{2-\alpha_\mathrm{e,inj}} / (2A_\mathrm{bol})$, that accounts for the fraction of total electron luminosity that could potentially be converted to a specific synchrotron luminosity at frequency $\nu$, given a magnetic field $B$ and electrons with Lorentz factor $\gamma_\mathrm{e}(\nu,B)$ (Eq.~\ref{eq: nu_synchr}). In addition, we introduce the calorimetric synchrotron fraction $\eta_{\mathrm{syn}}$, that quantifies the fraction of the available CR electron luminosity $\zeta_{\mathrm{bol}}L_\mathrm{e}$ that is actually converted to synchrotron emission, which will be discussed in more detail in Section~\ref{sec:testing electron calorimetry}.

Equations (\ref{eq:Q_e}) to (\ref{eq:FRC_L_nu2}) can be separably adapted to primary and secondary electrons. Hence, we arrive at an FRC that reads for the specific luminosity as a function of FIR luminosity for each CR electron population as
\begin{align}
\nu L_{\nu,\mathrm{prim}}\mathrm{(GHz)}\approx 1.0 \times 10^{-7}\,\left(\frac{\eta_{\mathrm{syn,prim}}}{0.1}\right)\left(\frac{\zeta_\mathrm{bol}}{0.02}\right)   L_{\mathrm{FIR}},
\label{eq:FRC_nu_L_prim(FIR)}
\end{align}
and
\begin{align}
\nu L_{\nu,\mathrm{sec}}\mathrm{(GHz)}\approx 1.3\times 10^{-8}\, \left(\frac{\eta_{\mathrm{cal,p}}}{0.8}\right) \left(\frac{\eta_{\mathrm{syn,sec}}}{0.1} \right) \left(\frac{\zeta_\mathrm{bol}}{0.001}\right) L_{\mathrm{FIR}},
\label{eq:FRC_nu_L_sec(FIR)}
\end{align}
where we use $\zeta_{\mathrm{SN}}=0.05$.
The 1.4~GHz radio luminosity as a function of SFR reads \citep[using again][]{1998Kennicutt},
\begin{align}
L_{1.4\mathrm{GHz,prim}}\approx 2.0\times 10^{27}\,\left(\frac{\eta_{\mathrm{syn,prim}}}{0.1}\right) \left(\frac{\zeta_\mathrm{bol}}{0.02}\right) \left(\frac{\dot{M}_{\star}}{\mathrm{M_{\odot}\,yr^{-1}}}\right),
\label{eq:FRC_L_prim(SFR)}
\end{align}
and
\begin{align}
L_{1.4\mathrm{GHz,sec}}\approx 2.6\times 10^{26}\,\left(\frac{\eta_{\mathrm{cal,p}}}{0.8}\right) \left(\frac{\eta_{\mathrm{syn,sec}}}{0.1}\right) \left(\frac{\zeta_\mathrm{bol}}{0.001} \right) \left(\frac{\dot{M}_{\star}}{\mathrm{M_{\odot}\,yr^{-1}}}\right),
\label{eq:FRC_L_sec(SFR)}
\end{align}
where we adopted $\gamma_\mathrm{e}=6\times10^{3}$, which is the typical Lorentz factor of CR electrons emitting synchrotron radiation at 1.4~GHz in magnetic fields of $2\,\umu\mathrm{G}$ (see Eq.~\ref{eq: nu_synchr}) and $\zeta_\mathrm{bol}(p_\mathrm{min}=1,\alpha_\mathrm{e,inj}=2.2,\gamma_\mathrm{e}=6\times10^{3})\approx 0.02$ for primary electrons, as well as $\zeta_\mathrm{bol}(p_\mathrm{min}=1,\alpha_\mathrm{e,inj}=2.7,\gamma_\mathrm{e}=6\times10^{3})\approx 0.001$ secondary electrons. This analytical estimate yields a synchrotron luminosity from secondary electrons that is a factor of $\sim10$ smaller than the luminosity arising from primary electrons, if the assumptions $\eta_{\mathrm{cal,p}}=0.8$, $\eta_{\mathrm{syn,prim}}=\eta_{\mathrm{syn,sec}}=0.1$ and $\zeta_{\mathrm{bol}}=0.02$ hold for primary electrons, whereas $\zeta_{\mathrm{bol}}=0.001$ for secondary electrons. We will discuss the choice of these parameters in the following.

The primary and secondary contributions to the total synchrotron luminosity $L_{1.4\,\mathrm{GHz}}$ in our simulations (Fig.~\ref{fig: FIR-Radio}, dark-gray symbols) are separately shown in the upper panels of Fig.~\ref{fig: FIR-Radio} with different colors, as indicated in the legend. The lower panels show the fraction of the primary and secondary luminosities, respectively. Whereas the primary luminosities fall short of the secondary emission in the `CR adv' model (right-hand panels), the radio luminosity is strongly dominated by primary electrons in our `CR diff' model (left-hand panels), with an increasing contribution from secondaries towards higher SFRs. This is due to the difference in both models in the parameters entering Eq.~(\ref{eq:FRC_L_sec(SFR)}). 

First, the calorimetric proton fraction $\eta_{\mathrm{cal,p}}$, representing the efficiency of secondary electron production due to hadronic interactions, has been shown in \citetalias{2021WerhahnII} to decrease with decreasing SFR in the `CR diff' model from 0.7 in starburst galaxies, down to 0.1 in our dwarf galaxies. In the `CR adv model', $\eta_{\mathrm{cal,p}}$ reaches values up to 0.8 and only typically varies by a factor of $\sim2$. The reduced efficiency of pion production due to CR diffusion can be understood by the fact that CR diffusion is another CR proton loss process that competes with hadronic losses and thus lowers the secondary radio luminosity. However, the difference of $\eta_{\mathrm{cal,p}}$ for star-forming galaxies with $\dot{M}_{\star} \gtrsim 1~\mathrm{M_\odot~yr^{-1}}$ is not significant enough to fully explain the discrepancy in $L_\mathrm{prim}$ vs. $L_\mathrm{sec}$ in our different CR transport models. 

A more relevant effect for explaining the sub-dominant role of secondary electrons for the total radio luminosity in the `CR diff' model is the effect of the steepening of the CR proton spectra due to energy-dependent diffusion. As an example, we show in Fig.~\ref{fig: CR-spectra} CR spectra of CR protons, primary and secondary electrons of a simulation with $M_{200}=10^{12}\,\mathrm{M_\odot}$ for both CR transport models at $t=1.3$~Gyr. The averaged CR spectra (shown with colours) are compared to momentum power-law spectra that are typical for the various transport and cooling processes (shown with grey-dashed lines), in which we account for the full proton/electron dispersion relation. This causes the mild downturn of the proton spectrum at energies $E_\rmn{kin}\lesssim 1$~GeV. We can clearly see the effect of Coulomb-cooling, which suppresses the spectra in comparison to the momentum power-law spectrum and which is stronger for CR protons in comparison to primary electrons, due to the dependence of Coulomb-cooling on $\beta=\varv/c$ (see Eq.~\eqref{eq: tau_coul} and \citetalias{2021WerhahnI}). 

At high energies, the CR proton spectrum stays flat in the `CR adv' model, i.e.\ $\alpha_\mathrm{p,adv}=\alpha_\mathrm{inj}=2.2$, whereas the spectral index approaches $\alpha_{\mathrm{p,diff}}=2.7$ in the `CR diff' model, which accounts for energy-dependent diffusion, i.e.\ $D\propto E^{0.5}$. Because the steady-state proton distribution determines the source function of secondary electrons, $\alpha_{\mathrm{sec-e,inj}}=\alpha_{\mathrm{p}}$, the latter also exhibits a steeper steady-state spectrum in the model accounting for energy-dependent diffusion losses after accounting for the radiative steepening of the electron spectra due to IC and synchrotron interactions, $\alpha_\mathrm{sec-e}= \alpha_{\mathrm{p,diff}}+1= 3.7$. By contrast, the steady-state spectrum of primary electrons has a similar spectral shape in both models, with $\alpha_\mathrm{prim-e}= 3.2$ at high energies, where IC and synchrotron cooling dominate. Note that the spectra in Fig.~\ref{fig: CR-spectra} are averaged over the radii that include 99 per cent of the total radio luminosity in both snapshots, respectively, and a characteristic scale-height of the magnetic field (as indicated in the legend) so that we show representative CR spectra for explaining the radio synchrotron emission at 1.4~GHz.

\begin{figure}
\includegraphics[]{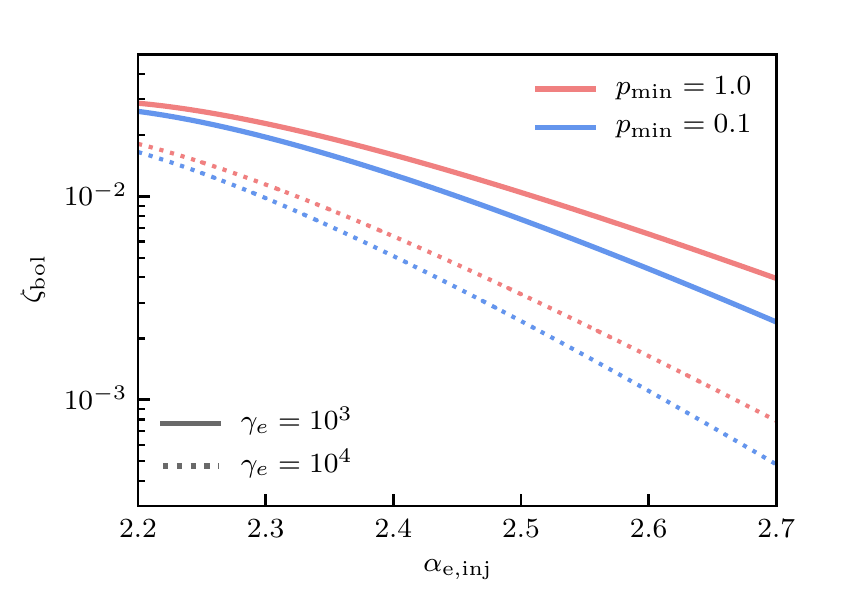}
\caption{We show the bolometric electron fraction $\zeta_\mathrm{bol}$ as implicitely defined in Eq.~(\ref{eq:FRC_L_nu2}). It quantifies the fraction of the total electron luminosity (from a CR electron source function approximated by a power law with injection spectral index $\alpha_\mathrm{e,inj}$ and low-momentum cutoff $p_{\mathrm{min}}$, see Eqs.~\ref{eq:Q_e} and \ref{eq:L_e1}) that is available for synchrotron emission at a frequency $\nu(\gamma_\mathrm{e},B)$. For instance, in an ambient magnetic field of $B=2~\umu\mathrm{G}$, electrons with a Lorentz factor $\gamma_\mathrm{e}=10^{4}$ radiate synchrotron emission at GHz frequencies (see Eq.~\ref{eq: nu_synchr}). }
\label{fig: zeta_bol}
\end{figure}

This effect of steeper secondary CR electron spectra is entering Eq.~(\ref{eq:FRC_L_sec(SFR)}) in the form of $\zeta_\mathrm{bol}$, that is a function of $p_\mathrm{min}$, $\gamma_\mathrm{e}(\nu,B)$ and $\alpha_\mathrm{e}$ (see Eq.~\ref{eq:FRC_L_nu2}). We exemplify these dependencies in Fig.~\ref{fig: zeta_bol}, that shows the strong decrease of $\zeta_\mathrm{bol}$ with increasing spectral index $\alpha_\mathrm{e}$ of the CR electron source function, which affects the available luminosity of primary and secondary electrons, that can be converted to radio synchrotron emission. As expected, steeper electron spectra imply a smaller amount of electrons available for synchrotron emission at a fixed frequency. While the low-momentum cut-off of the CR electron spectra does not have a significant impact on $\zeta_\mathrm{bol}$, the electrons' Lorentz factor $\gamma_\mathrm{e}$ naturally becomes increasingly more relevant for increasing spectral indices. As a result, $\zeta_\mathrm{bol}$ is affected by the $\nu_\mathrm{c}$-effect.

Consequently, we identify two main effects that can potentially be responsible for suppressing the radio luminosity via $\zeta_\mathrm{bol}$: 
(i) the steeper secondary CR electron source functions with the limiting value of $\alpha_\mathrm{sec-e,inj}=2.7$ due to CR diffusion decrease $L_{1.4\mathrm{GHz,sec}}$ (in comparison to primaries with $\alpha_{\mathrm{prim-e, inj}}=2.2$ in both models and also secondaries with $\alpha_{\mathrm{sec-e, inj}}=2.2$ in the `CR adv' model), and 
(ii) the $\nu_\mathrm{c}$-effect, i.e., synchrotron emission at a fixed frequency in a lower magnetic field strength is generated by higher-energetic electrons, which are less abundant. This implies a smaller value of $\zeta_\mathrm{bol}$, which leads to a lower observed radio luminosity. The second effect is subdominant for primary CR electrons in both CR transport models and for secondary CR electrons in the `CR adv' model because $\zeta_\mathrm{bol}$ does not significantly depend on $\gamma_\mathrm{e}$ for $\alpha_\mathrm{e,inj}=2.2$. However, with increasingly larger $\alpha_\mathrm{e,inj}$, $\zeta_\mathrm{bol}$ is reduced by up to a factor of 5, when we increase $\gamma_\mathrm{e}$ by an order of magnitude.
This is in fact the case within our simulations: As pointed out by \cite{2021Pfrommer}, our simulations exhibit saturated magnetic field strengths of $0.1~\umu$G to $14~\umu$G for SFRs of $0.01$ to $30~\mathrm{M_\odot}~\mathrm{yr}^{-1}$ in our `CR diff' model. This implies typical Lorentz factors of electrons emitting at 1.4~GHz that range from $\gamma_\mathrm{e}\approx3.5\times 10^4$ to $\gamma_\mathrm{e}\approx 3.5\times10^3$. As a result the $\nu_\mathrm{c}$-effect indeed plays an important role and is particularly strong for secondary electrons:
as we move from starburst to dwarf galaxies the saturated magnetic field strengths decrease and imply increasing Lorentz factors of CR electrons, which diminishes the total radio luminosity (if it is observed at a fixed frequency). But in the case of our `CR diff' model, diffusive losses become increasingly more important towards lower star-forming galaxies (\citetalias{2021WerhahnII}), which steepens $\alpha_\mathrm{p}=\alpha_\mathrm{inj,sec.e}$ and hence, the secondary contribution to the total radio luminosity experiences an additional suppression, which explains the decreasing fraction of secondary radio luminosity with decreasing SFR, as shown in the lower left-hand panel in Fig.~\ref{fig: FIR-Radio} for our `CR diff' simulations.

\subsection{Testing electron calorimetry \label{sec:testing electron calorimetry}}

\begin{figure*}
\begin{centering}
\includegraphics[scale=1]{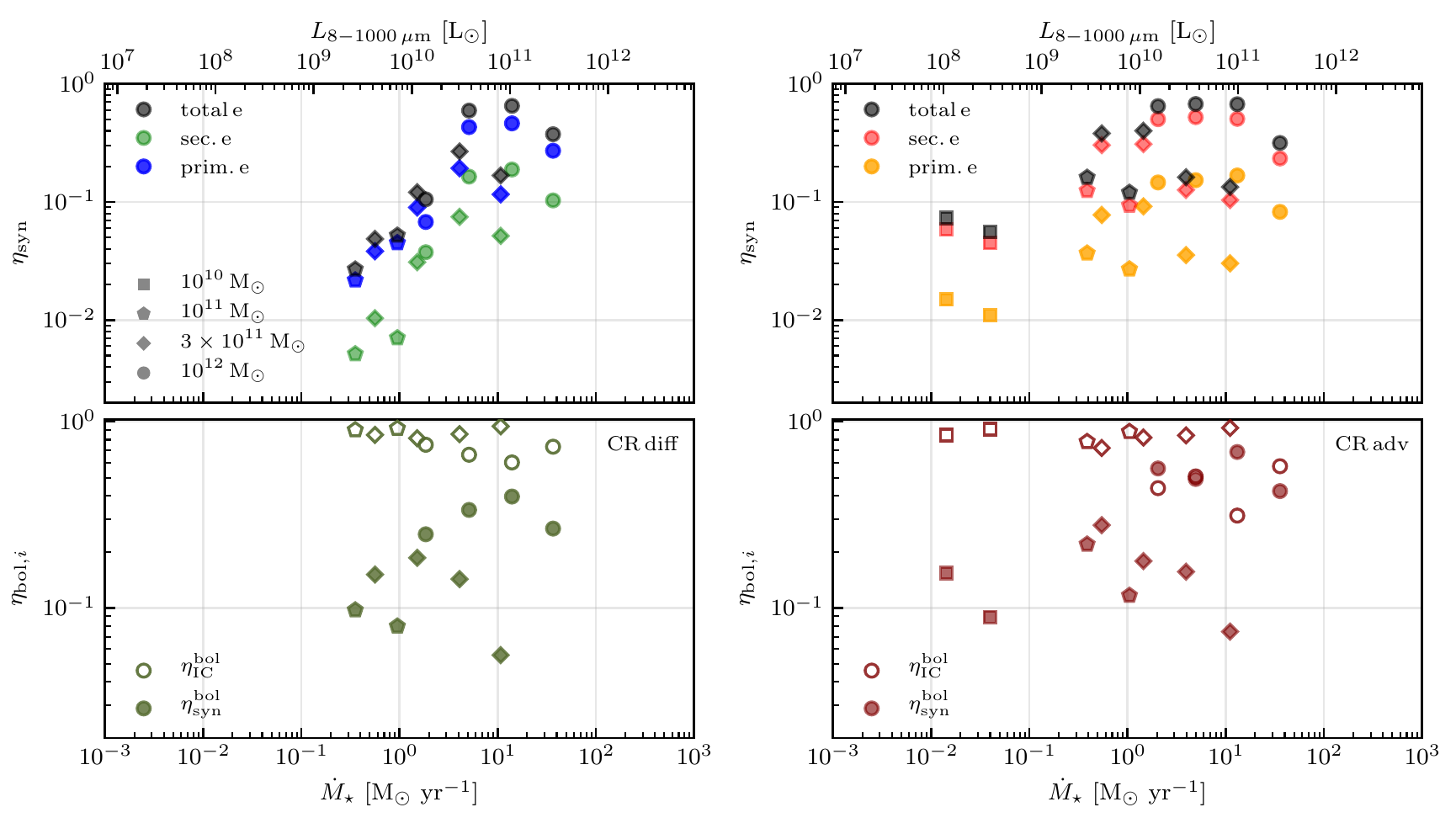}
\par\end{centering}
\caption{Top panels: we show the the calorimetric synchrotron fraction (see Eq.~\ref{eq: eta-syn}) for our CR diffusion (left-hand panel) and advection model (right-hand panel) as a function of SFR. In addition to the total fraction including primary and secondary electrons (black), we also show the fractions for primaries and secondaries separately (colour coded as indicated in the legend). Bottom panels: here, we show the bolometric fractions of synchrotron (filled symbols) and IC emission (open symbols) defined in Eqs.~\eqref{eq:eta_bol_syn} and \eqref{eq:eta_bol_IC} which add up to unity by construction. The different symbols correspond to simulations with different halo masses as indicated in the top left-hand panel.}
\label{fig: calorimetric_fraction}
\end{figure*}

In order to quantify electron calorimetry of our simulated galaxies as a function of SFR and halo mass, we first define the calorimetric synchrotron fraction following Eq.~(\ref{eq:FRC_L_nu2}) as

\begin{align}
\eta_{\mathrm{syn,prim/sec}}=\frac{\nu L_{\nu,\mathrm{prim/sec}}}{\zeta_\mathrm{bol} L_\mathrm{e}}
=\frac{ \sum_i \nu L_{\nu,\mathrm{prim/sec},i}}{\sum_i\zeta_{\mathrm{bol},i} L_{\mathrm{e},i}},
\label{eq: eta-syn}
\end{align}
where in each cell $i$, we define the fraction of the total CR electron luminosity that radiates synchrotron emission at a frequency $\nu$, given a magnetic field $B_i$ within a cell $i$,
\begin{align}
\zeta_{\mathrm{bol},i} = \frac{\left[\gamma_{\mathrm{e}}(B_i,\nu)\right]^{2-\alpha_\mathrm{e,inj}}}{2 A_\mathrm{bol}},
\end{align}
and $L_\mathrm{e}=L_\mathrm{prim-e} + L_\mathrm{sec-e}$ denotes the electron luminosity. For simplicity, the calorimetric synchrotron fractions of primary and secondary electrons $\eta_{\mathrm{syn, prim/sec}}$, are defined in such a way that they add up to a total calorimetric synchrotron fraction
\begin{align}
\eta_{\mathrm{syn}}=\eta_{\mathrm{syn,prim}} + \eta_{\mathrm{syn,sec}} = \frac{\nu L_\nu}{\zeta_\mathrm{bol} L_\mathrm{e}}.
\label{eq: eta-syn-total}
\end{align}
To fulfill this condition of additivity, we assume $\alpha_\mathrm{e,inj}=2.2$ in all cells for both primary and secondary electrons. This implies that $\eta_{\mathrm{syn,sec}}$ represents in our definition a lower limit for the calorimetric synchrotron fraction of secondary electrons, since they can exhibit steeper injected spectral indices $\alpha_\mathrm{sec.e,inj}>2.2$, if energy dependent CR diffusion is included, as discussed in Section~\ref{sec:secondary vs. primary synchr. emission}.

To calculate these calorimetric synchrotron fractions of our simulations, we estimate the available proton luminosity from the SFR using Eq.~(\ref{eq: L_p1}) in each cell,\footnote{Note that this implies that we only sum over cells with a SFR $\dot{M}_\star>0$.} and use this to compute the electron luminosities from Eq.~(\ref{eq: L_e_prim,e}) and (\ref{eq: L_e_sec,e}).
In the upper panels of Fig.~\ref{fig: calorimetric_fraction}, we show the calorimetric synchrotron fractions of our simulated galaxies (the same snapshots as shown in Fig.~\ref{fig: FIR-Radio}) for primary and secondary electrons, as well as the total calorimetric fraction. 
These range from 0.03 to 0.65 in our `CR diff' models and from 0.06 to 0.67 in our `CR adv' models. This partly explains the large scatter that we obtain in our FRC in Fig.~\ref{fig: FIR-Radio}. In particular, there is a difference in the calorimetric synchrotron fraction for different halo masses at the same SFR, that leads to a spread in radio luminosities. This is a result of the growth of the magnetic dynamo that continues to amplify magnetic fields at large disc radii of the galaxies, after saturation of the small-scale magnetic dynamo \citep{2021Pfrommer}.

Furthermore, we observe a global trend towards higher calorimetric fractions with higher SFRs, which is particularly strong in the `CR diff' models. It remains to be seen whether an improved model for the ISM in global (dwarf) galaxy simulations with explicit SN-driven turbulence \citep{2018Semenov,2021Gutcke} in a cosmological setting can further amplify the magnetic field in the outskirts of the disc via a small-scale dynamo. This would increase the values of $\eta_\rmn{syn}$ at low SFRs in dwarfs over what is found in our models, where fast CR diffusion in the `CR diff' model in low-mass halos efficiently quenches the small-scale dynamo and hence suppresses the calorimetric synchrotron fractions.

So far, we only considered the fraction of the injected electron luminosity that can potentially be converted to synchrotron emission at a fixed frequency, i.e.\ we only selected electrons with a certain Lorentz factor $\gamma_\mathrm{e}(\nu, B)$ that emits into a given frequency window. But in addition to that, we are also interested in comparing the total (bolometric) amount of energy that is radiated via synchrotron emission in comparison to IC scattering.
To extend these considerations over the whole energy range, we define bolometric fractions of synchrotron and IC emission according to
\begin{align}
\eta_{\mathrm{syn}}^{\mathrm{bol}} = \frac{\sum_i B_i^2V_i}{\sum_i (B_i^2+B_{\mathrm{ph},i}^2)V_i}
\label{eq:eta_bol_syn}
\end{align}
and
\begin{align}
\eta_{\mathrm{IC}}^{\mathrm{bol}} = \frac{\sum_i B_{\mathrm{ph},i}^2V_i}{\sum_i (B_i^2+B_{\mathrm{ph},i}^2)V_i}.
\label{eq:eta_bol_IC}
\end{align}
These definitions are derived from the synchrotron and IC loss rates, which depend on the energy density of the magnetic field $\varepsilon_B\propto B^2$ and the photon energy density $\varepsilon_\mathrm{ph}\propto B_\mathrm{ph}^2$, respectively (see Eqs.~\ref{eq: tau_syn} and \ref{eq: tau_IC}). 
Per definition, $\eta_{\mathrm{IC}}^{\mathrm{bol}}+\eta_{\mathrm{syn}}^{\mathrm{bol}}=1$, and hence, these fractions provide a measure of the fraction of the total electron luminosity that is lost to IC or synchrotron emission (at all possible frequencies) if we only consider these two radiative loss processes. To be consistent with the calculation of $\eta_{\mathrm{syn}}$, we again only sum over all cells with SFRs $\dot{M}_\star>0$ and show the resulting bolometric fractions of IC and synchrotron emission in the lower panels of Fig.~\ref{fig: calorimetric_fraction}.
In almost all analysed snapshots, we find that $\eta_{\mathrm{IC}}^{\mathrm{bol}} > \eta_{\mathrm{syn}}^{\mathrm{bol}}$, which implies that IC losses are usually dominating synchrotron losses for CR electrons. For our small halos with $M_{200}=10^{10}\,\mathrm{M_\odot}$ and $M_{200}=10^{11}\,\mathrm{M_\odot}$, this is not surprising: \cite{2021Pfrommer} found that the magnetic field in these small halos saturate below the energy density of the CMB and hence, IC-cooling via scattering off of the CMB alone already dominates over synchrotron cooling. There are only three cases for which $\eta_{\mathrm{IC}}^{\mathrm{bol}} \leq \eta_{\mathrm{syn}}^{\mathrm{bol}}$, which represent highly star-forming galaxies with $M_{200}=10^{12}\,\mathrm{M_\odot}$ in the `CR adv' model. They saturate at the highest magnetic field strengths in comparison to all other shown simulations \citep{2021Pfrommer} and hence, result in the highest bolometric fractions of synchrotron emission, that manage to dominate over IC emission.

\subsection{`Conspiracy' at high gas densities} \label{sec:conspiracy at high densities}

\begin{figure*}
\begin{centering}
\includegraphics[scale=1]{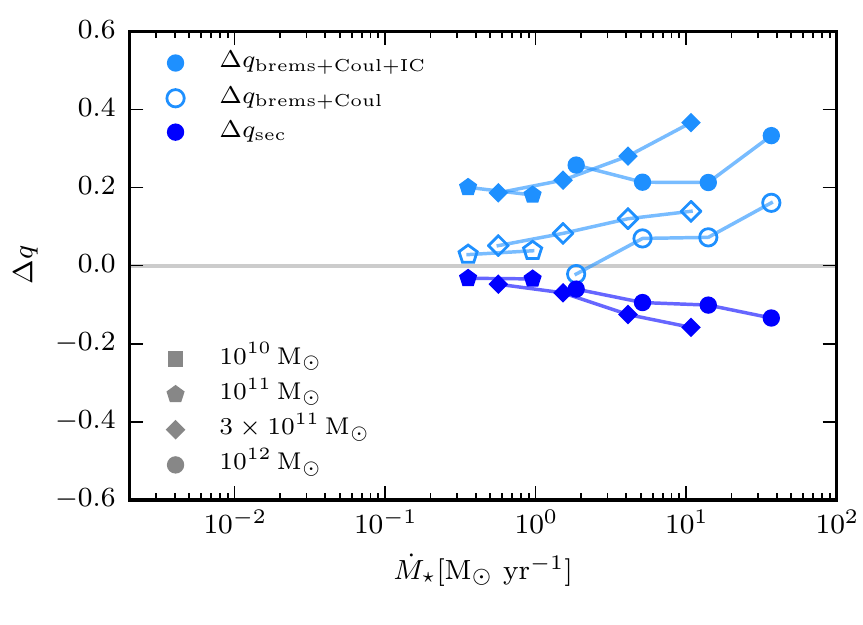}\includegraphics[scale=1]{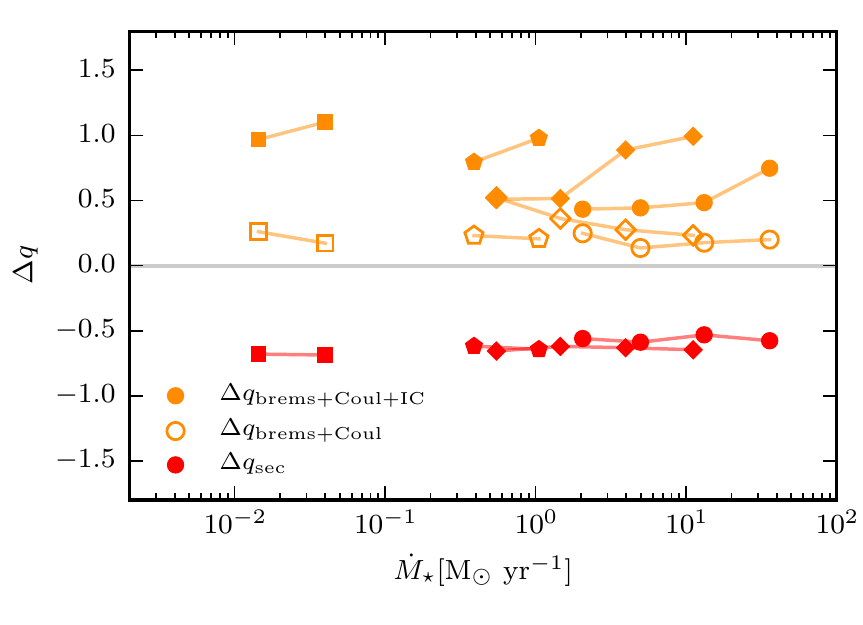}
\par\end{centering}
\caption{We show the change in the FIR-to-radio luminosity ratio, quantified by $q$ (see Eq.~\ref{eq: q}), for our `CR diff' (left-hand panel) and `CR adv' models (right-hand panel). Here, we study how individual cooling processes modify the FRC: the effect of neglecting bremsstrahlung and Coulomb losses, that both depend on gas density, is shown via $\Delta q_\mathrm{brems+Coul}$ (Eq. ~\ref{eq: delta_q_bremsCoul}) in light blue (orange) with open symbols in our `CR diff' (`CR adv') models, whereas the effect of additionally neglecting IC losses, $\Delta q_\mathrm{brems+Coul+IC}$, is shown by the corresponding full symbols. Changes in $q$ that result from disregarding the secondary radio emission are denoted by $\Delta q_{\mathrm{sec}}$ (Eq.~\ref{eq: delta_q_sec}) and are shown in dark blue (red). Different symbols correspond to different halo masses as indicated in the left-hand panel. Note that both panels have different ranges on their vertical axes. }
\label{fig: delta_q}
\end{figure*}

We found in Section~\ref{sec:secondary vs. primary synchr. emission}, that there is an increasing contribution of secondary emission to the total radio luminosity toward larger SFRs in our `CR diff' model. If this trend holds, the question arises of how to maintain the almost linear behaviour of the FRC.
As proposed by \citet{2010Lacki}, in addition to the larger contribution of secondary electrons to the radio luminosity in high-density starburst galaxies, one would also expect more losses due to Coulomb interactions and bremsstrahlung emission due to their dependence on gas density, see Eqs.~(\ref{eq: tau_coul}) and (\ref{eq: tau_brems}). Furthermore, IC losses are expected to increase with higher SFRs due to the higher photon energy densities from young stellar populations. Consequently, in order to maintain an almost linear FRC, those effects have to `conspire' to (approximately) cancel each other \citep{2010Lacki}.

To quantify deviations from the FRC and to test the relative contributions of the proposed effects, we utilise the ratio of total IR (TIR) to radio synchrotron emission as defined by \cite{1985Helou}:
\begin{align}
q=\log_{10}\left(\frac{L_{\mathrm{TIR}}}{L_{1.4\,\mathrm{GHz}}} \right) - 3.67,
\label{eq: q}
\end{align}
where the TIR luminosity is given by $L_{\mathrm{TIR}}\approx 1.75 L_{\mathrm{FIR}}$ \citep{2000Calzetti}.
We define the deviation from a given FRC via the difference $\Delta q$ that results from comparing the full and a modified FRC for which we exclude different contributions to the total synchrotron emission. This enables us to study the impact of individual processes in modifying the slope of the FRC and whether these processes pose a challenge to the observed quasi-linearity of the FRC.

First, we assess the effect of neglecting the contribution of secondary electrons to the total radio luminosity $L_{1.4\,\mathrm{GHz}}$, and denote the corresponding parameter with $q_{\mathrm{no\ sec}}$. This is expected to be larger than $q$, where primary and secondary contributions are included. Hence, the difference 
\begin{align}
\Delta q_{\mathrm{sec}}= q - q_{\mathrm{no\ sec}}\leq 0
\label{eq: delta_q_sec}
\end{align}
enables us to quantify the contribution of secondary radio emission to the FRC.

Second, we consider the case of disregarding losses due to bremsstrahlung and Coulomb interactions to infer their effect on the radio emission and define the resulting difference as
\begin{align}
\Delta q_{\mathrm{brems+Coul}} = q - q_{\mathrm{no\  brems+Coul}}.
\label{eq: delta_q_bremsCoul}
\end{align}
To calculate $q_{\mathrm{no\  brems+Coul}}$, we compute the radio luminosity in Eq.~(\ref{eq: q}) from steady-state spectra that we obtain by setting $b_{\mathrm{Coul}}=b_{\mathrm{brems}}=0$. If these losses are significant in shaping the steady-state distribution, we expect $q_{\mathrm{no\,brems+Coul}}\leq q$ because fewer losses due to processes other than synchrotron emission potentially give rise to a larger radio luminosity. Similarly, we calculate the change in $q$ from neglecting all non-synchrotron cooling processes, i.e.\ bremsstrahlung, Coulomb and IC cooling, which is denoted by 
\begin{align}
\Delta q_{\mathrm{brems+Coul+IC}} = q - q_{\mathrm{no\ brems+Coul+IC}}.
\label{eq: delta_q_bremsCoulIC}
\end{align}

Figure~\ref{fig: delta_q} shows the deviation from a linear FRC when excluding secondary radio emission or bremsstrahlung and Coulomb losses. In our `CR diff' runs, both effects become relevant in galaxies with SFRs $\dot{M}_\star \gtrsim 1~\mathrm{M_\odot\,yr^{-1}}$.
In these star-forming and star-bursting galaxies, the missing radio emission from excluding secondary CR electrons is compensated by the additional radio emission, resulting from the neglect of bremsstrahlung and Coulomb losses. As a consequence, the deviation from calorimetry that would be expected in starbursts, due to the increasing relevance of bremsstrahlung and Coulomb interactions in these high-density galaxies, seems to be counterbalanced by including the contribution of secondary radio emission. 

IC losses elevate $q$ across all SFRs and show an almost constant shift of $q$ for our `CR diff model', indicating that the IC emission does not significantly modify the slope of the FRC in this model. While IC losses become increasingly more relevant toward high SFRs for every individual galaxy simulation in our `CR adv' model, on average the distribution of $\Delta q_{\mathrm{brems+Coul+IC}}$ is an elevated version of $\Delta q_{\mathrm{brems+Coul}}$ with an increased scatter. Hence, in spite of most of CR electrons lose their energy through IC interactions (Fig.~\ref{fig: calorimetric_fraction}), this shows that the subdominant synchrotron cooling process nevertheless appears to be a quasi-calorimetric tracer of the SFR, implying an almost linear slope of the FRC. Note that the overall change in $q$ is comparably small in the `CR diff' model and $\Delta q \leq 0.37$ for all considered effects. For a mean observed $q$ of 2.21 in starburst galaxies \citep{1985Helou}, this is at most a 17 per cent effect.

In contrast to our `CR diff' model, galaxies of our `CR adv' model (right-hand panel in Fig.\ref{fig: delta_q}) exhibit a contribution of secondaries that is almost independent of SFR. 
This is because neglecting diffusion losses increases the importance of hadronic losses, which are faster than escape losses $\tau_{\pi}\ll \tau_\mathrm{esc}$. As a result, the ratio of primaries to secondaries does not depend on the gas density or the SFR, but only depends on $K_\mathrm{ep}^{\mathrm{inj}}$ and $\alpha_\mathrm{p}$ (see Eq.~\ref{eq:N_e_prim/N_e_sec}), and is hence approximately constant across the range of SFRs probed by our set of simulations that only accounts for CR advection. Furthermore, as discussed in Section~\ref{sec:secondary vs. primary synchr. emission},  the $\nu_\mathrm{c}$-effect can only significantly affect secondary electrons in the model accounting for energy dependent CR diffusion, which results in an SFR-dependence of the secondary contribution to the total radio synchrotron emission and hence explains the decrease of $\Delta q_\mathrm{sec}$ with increasing SFR in the `CR diff' model.

\begin{table*}
 \caption{This table summarises the properties of our simulated galaxies, that resemble the observed galaxies NGC~253, M82 and NGC~2146 in terms of their SFRs and total radio luminosities. The observed radio luminosity is derived from the observed flux density at the frequency bin that is closest to $\nu=1.4\,\mathrm{GHz}$ using the data shown in Figs.~\ref{fig: radio-spectra NGC 253 and M82} and \ref{fig: radio-spectrum NGC2146} and from the distances summarised below. We use the simulated radio luminosities with the observationally determined inclination angle $\phi$ and adapt $\delta=0.3$. The central radius $R_{\mathrm{central}}$ and the constant fraction $\xi_\mathrm{e}$ of the electron density provided in our pressurised ISM model \citep{2003SpringelHernquist} are parameters used to construct the radio emission spectra from the simulations, including thermal free-free emission and absorption. }
 \label{Table-Galaxies}
 \begin{threeparttable}[t]
 \begin{tabular}{lccccccc}
  \hline
  Galaxy & SFR (obs.\tnote{1} /sim.) & Distance\tnote{2} & $L_{1.4\,\mathrm{GHz}}$ (obs./sim.) & $R_\mathrm{central}$(obs./sim.)&  $\phi$(obs.\tnote{3} /sim.) & $\xi_\mathrm{e}$ & Simulation\\
    & $[\mathrm{M}_{\odot}~\mathrm{yr}^{-1}]$&$\mathrm{[Mpc]}$ &$[\mathrm{erg~s~Hz}^{-1}]$& $\mathrm{[kpc]}$ & & & $M_{200}\,[\mathrm{M_{\odot}}],~t\,[\mathrm{Gyr}]$ \\
  \hline
    \hline
   NGC 253  &   $5.03 \pm 0.76$    &  $3.3$ & $7.67 \times 10^{28}$  & 0.15 - 0.25  & $74^{\circ}$& - & - \\
                     &   $4.110$                               &  - & $3.87\times 10^{28}$ & 0.15 & $74^{\circ}$  & 0.8 & $M_{200}=3\times 10^{11}$, $t=1.1$\\ 
    \hline
   M82\tnote{4} &  $10.4\pm 1.6$& $3.7$  &  $1.21\times 10^{29}$ & 0.45  &$80^{\circ}$ &-& - \\
                      &  $6.457$                                &  -  &  $ 1.62\times 10^{29}$   & 1.50    & $80^{\circ}$  &1.0 & $M_{200}=10^{12}$, $t=2.3$ \\ 
   \hline
   NGC 2146 & $14.0\pm 0.5$       & 15.2 & $3.02\times 10^{29}$ & -  & $63^{\circ}$ &-&- \\ 
                      & $25.520$                               &  -  & $5.55\times 10^{29}$ & 0.30  & $63^{\circ}$  &0.5 & $M_{200}=10^{12}$, $t=0.7$ \\
  \hline
 \end{tabular}
 \begin{tablenotes}
      \item[1] \citet{2020Kornecki} 
      \item[2] NGC~253: \cite{2005Mouhcine}; M82: \cite{2015Vacca}; NGC~2146: \cite{2004Gao}. 
      \item[3] NGC~253: \cite{2014Iodice}; M82: \cite{1963LyndsSandage}, \cite{1995McKeith}; NGC~2146: \cite{1999DellaCeca}. 
      \item[4] This snapshot is shown in Figs.~\ref{fig: tau}, \ref{fig: maps-properties}, \ref{fig: maps free-free}, \ref{fig: maps timescales} and \ref{fig: synchr_alpha_maps}.
   \end{tablenotes}
 \end{threeparttable}
\end{table*}

\section{Radio spectra} \label{sec:radio spectra}

As pointed out by \cite{2006Thompson}, the calorimetric assumption must hold for starburst galaxies like Arp~220. Here, the large photon energy density implies a short IC cooling time scale: 
\begin{align}
    \tau_\mathrm{IC}\sim 4\times10^3\,\mathrm{yr}\,
    \left(\frac{\varepsilon_\rmn{ph}}{10^{-6}~\rmn{erg~cm}^{-3}}\right)^{-1}\,
    \left(\frac{B}{3~\rmn{mG}}\right)^{1/2}.
\end{align}
In order for escape losses to be faster, unphysical wind velocities of order $\sim 20,000\,\mathrm{km\,s^{-1}}$ would be required for CR electrons to be advected from the compact star-bursting region of size $\sim100$~pc. Consequently, losses due to IC emission must be larger than escape losses and we would expect steep radio spectra. This is because an IC-cooled steady-state CR electron spectrum with $\alpha_\mathrm{e}=3.2$ implies a synchrotron spectral index of $\alpha_{\nu}=(\alpha_\mathrm{e}-1)/2=1.1$, which is larger than the observed spectral indices of $\sim 0.5$ to $0.8$. The same argument holds, if synchrotron losses dominate the cooling of CR electrons, due to the identical energy dependence of synchrotron and IC losses.

In order to solve this tension, we analyse in the following three possible mechanisms that could be responsible for flattening the radio spectra.
First, a non-negligible contribution of thermal free-free emission to the radio emission could have an influence on the radio spectral shape. 
Second, because bremsstrahlung losses have a weaker energy dependence, i.e.\ $b_{\mathrm{brems}} \propto E \ln E$ (Eq.~\ref{eq: tau_brems}), in comparison to IC and synchrotron losses with $b_{\mathrm{IC/syn}} \propto E^2$, they could flatten the electron spectra if they were faster than IC and synchrotron losses, i.e.\ $\tau_{\mathrm{brems}}<(\tau_{\mathrm{IC}}^{-1}+\tau_{\mathrm{syn}}^{-1})^{-1}$. Similarly, Coulomb cooling could affect the radio spectral shape because $b_\mathrm{Coul}\approx \mathrm{const.}$ for $\gamma_\mathrm{e}\gg 1$.
A third possibility for flattening the electron and hence the emitted synchrotron spectra is advection. Because advection losses do not depend on energy, the injected electron spectral index of 2.2 would be maintained in the regions where those losses are dominant. 
We test these possibilities in the following sections. 

So far,  in our fiducial model we adopted an energy dependence of the diffusion coefficient $D\propto E^{\delta}$ where $\delta=0.5$. In order to be consistent with our findings in \citetalias{2021WerhahnII}, where we found that a shallower dependence of $\delta=0.3$ enables better matches to the observed gamma-ray spectra of NGC~253, M82 and NGC~2146, we adopt in the following $\delta=0.3$ for the calculation of the radio spectra.\footnote{The energy dependence of the diffusion coefficient does not significantly change the radio spectra of these galaxies because radiative losses of CR electrons beat their diffusive losses which is different for CR protons that do not suffer radiative losses. For the considered snapshots, the effect of changing $\delta=0.5\to0.3$ only changes $L_{1.4\,\mathrm{GHz}}$ by less than 8 per cent.}

\subsection{Thermal and non-thermal radio emission\label{sec: radio spectra: thermal and non-thermal radio emission}}

\begin{figure*}
\begin{centering}
\includegraphics[scale=1]{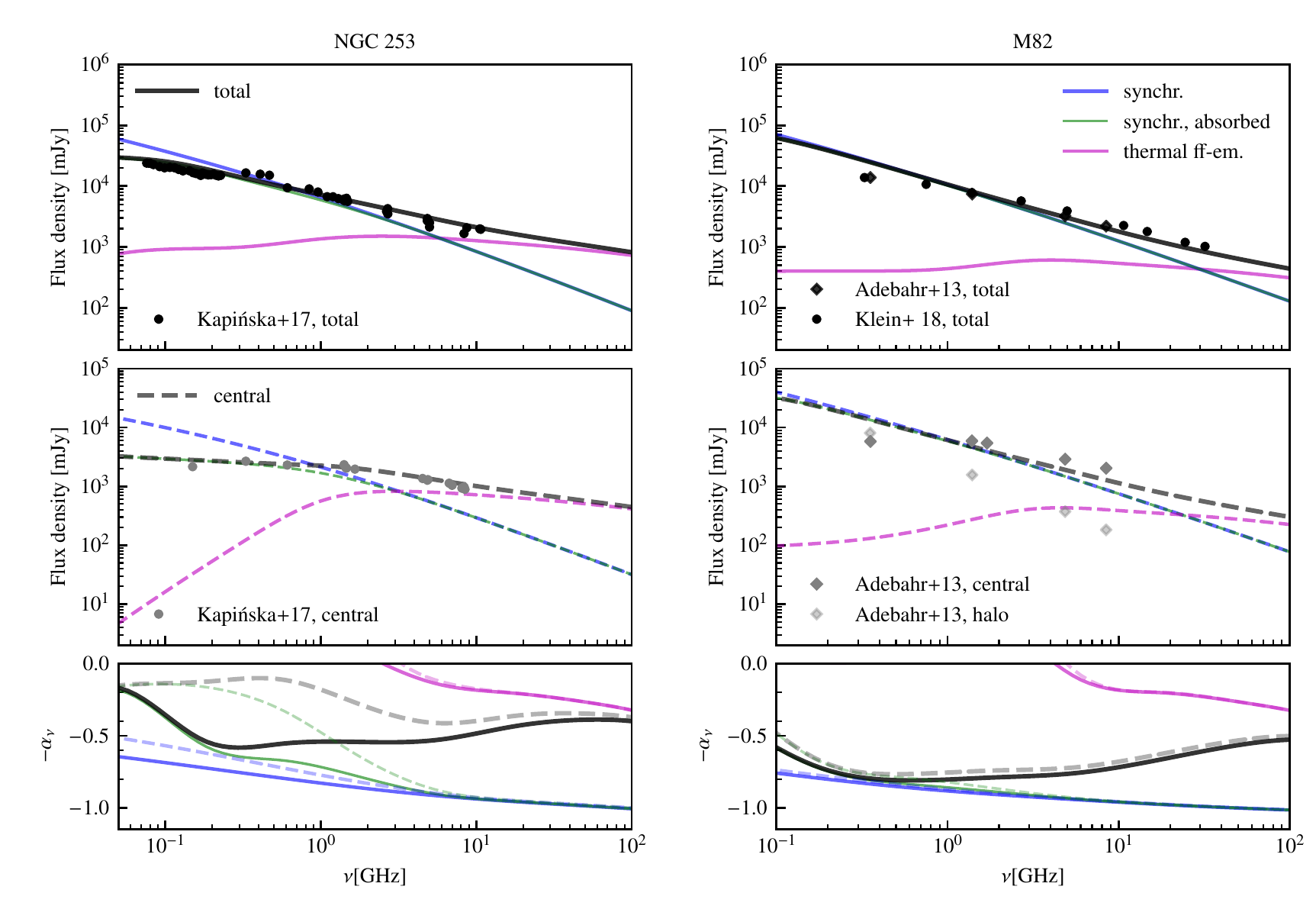}
\par\end{centering}
\caption{Spectra of synchrotron emission (blue lines) and thermal free-free emission (magenta lines) of our simulations that are similar to NGC~253 and M82 in terms of SFR and total radio luminosity (see Table~\ref{Table-Galaxies}). The spectra are computed for inclination angles $\phi=74^\circ$ for NGC~253 \citep{2014Iodice} and $\phi=80^\circ$ for M82 \citep{1963LyndsSandage,1995McKeith}. 
The upper panels show the spectra for the whole galaxies, whereas for the middle panels, we cut out a region around the galactic center with a radius $R_{\mathrm{central}}$ (see Table~\ref{Table-Galaxies}; dashed lines), resembling the observed central regions, respectively. To allow a comparison with the spectral shapes of the observed spectra, the simulated spectra are re-normalised by $L_{1.4\mathrm{GHz,obs}}/L_{1.4\mathrm{GHz,sim}}$ (see Table~\ref{Table-Galaxies}). Thermal free-free absorption affects the synchrotron spectra at low frequencies (green lines) and is strongest for the central spectra, due to the high central gas density. For NGC~253 and M82, we plot the total (black points) and central (grey points) observed radio spectrum by \citet{2017Kapinska}, \citet{2018Klein} and \citet{2013Adebahr_M82}, as indicated in the legend. The light grey points in the middle right-hand panel correspond to the spectrum of the halo, i.e.\ the difference between the total and central spectrum, which matches the spectral shape of our synchrotron component. Bottom panels: the solid (dashed) lines show the spectral indices of our total (central) simulated radio spectra for synchrotron, thermal free-free and total emission, respectively. For a variation of $R_\mathrm{central}$, see App.~\ref{app: R_central}. }
\label{fig: radio-spectra NGC 253 and M82}
\end{figure*}

\begin{figure}
\begin{centering}
\includegraphics[scale=1]{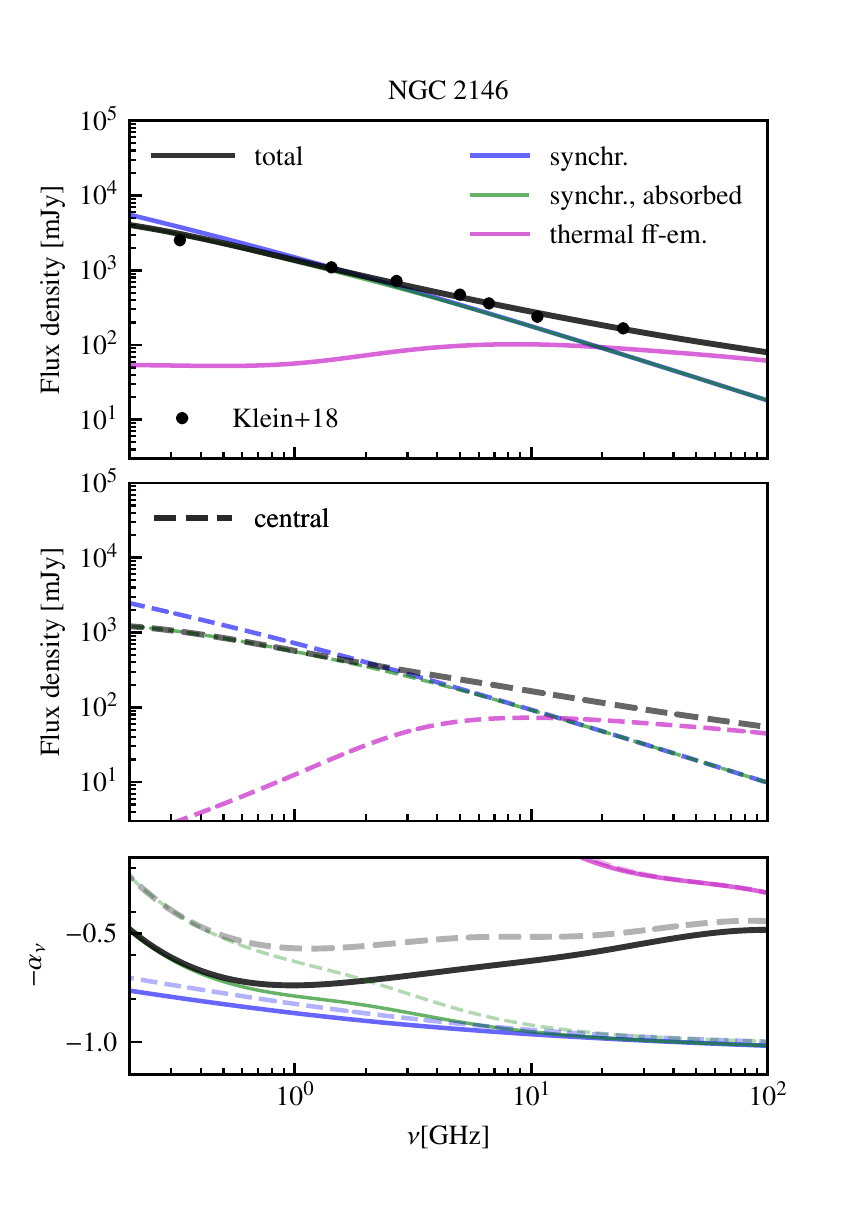}
\par\end{centering}
 \caption{Top panel: we compare the observed total radio spectrum of  NGC~2164 \citep{2018Klein} to simulated spectra of synchrotron and thermal free-free emission of a simulation that matches NGC~2164 in terms of SFR and total radio luminosity (see Table~\ref{Table-Galaxies}). Here, we adopt an inclination angle of $\phi=63^{\circ}$ \citep{1999DellaCeca}. The middle panel shows the spectra in the central region (within $R_\mathrm{central}=0.3\,\mathrm{kpc}$) while the bottom panel shows the spectral index of our simulated radio spectra for synchrotron, thermal free-free and total emission, respectively.}
\label{fig: radio-spectrum NGC2146}
\end{figure}

\begin{figure*}
\begin{centering}
\includegraphics[scale=1]{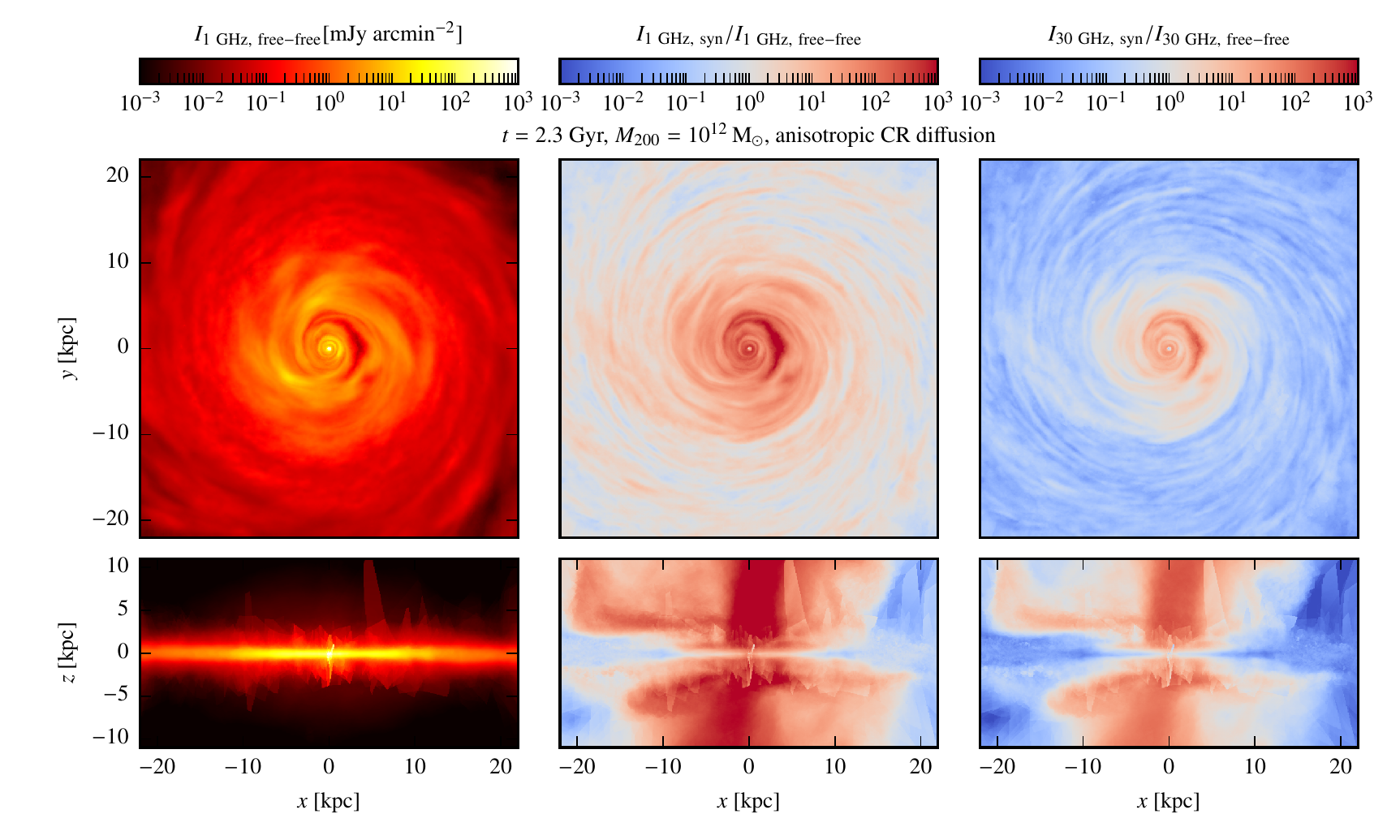}
\par\end{centering}
\caption{Face-on and edge-on maps of the projected thermal free-free emission at 1~GHz (left-hand panels) of our simulation with $M_{200}=10^{12}\,\mathrm{M_\odot}$ at $t=2.3$~Gyr. The middle and right-hand panels show the ratio of the synchrotron emission to the thermal free-free emission at 1 and 30~GHz, respectively.	}
\label{fig: maps free-free}
\end{figure*}

Figures~\ref{fig: radio-spectra NGC 253 and M82} and \ref{fig: radio-spectrum NGC2146} show the radio spectra of the three star-forming galaxies NGC~253, M82 and NGC~2146. We compare observations by \cite{2017Kapinska}, \citet{2013Adebahr_M82} and \citet{2018Klein} to radio spectra derived from our simulated galaxies that resemble the observed ones in terms of their SFRs and radio luminosities, respectively (see Table~\ref{Table-Galaxies}).
We show the observed and simulated total spectra (top panels) and the spectra from the central regions (middle panels). 
The flux density is inferred from our modeling of the intensity $I_\nu$, as it would be observed from a galaxy with an inclination angle $\phi$ at a luminosity distance $d$ (see Eqs.~\ref{eq: I_nu rad. transfer} and \ref{eq: F_nu}).
Below the total and central radio spectra, we show the spectral indices of the differently modeled components that are computed via 
\begin{align}
\alpha_\nu = -\frac{\mathrm{d}\log F_\nu}{\mathrm{d}\log \nu}.
\end{align}
We find that the total radio synchrotron spectra exhibit spectral indices from 0.6 to 1.1 for radio frequencies ranging from 0.1 to 100 GHz. Those are significantly steeper than the observed radio spectra at frequencies larger than a few GHz. However, our modeling of the thermal free-free emission (see Section~\ref{sec: free-free-emission and absorption}) is able to flatten the total spectra so that they approach the observed values also at frequencies $\nu\gtrsim1$~GHz.

The spectrum of the central region of a simulated galaxy is obtained by computing the intensity $I_\nu(\mathbfit{r}_\perp)$ within a circle around the center with the radius $R_\mathrm{central}$ (see Table~\ref{Table-Galaxies}) that approximately corresponds to the radius of the observed central region. Only for M82, we adapt a larger radius to match the total observed flux from the central region with our model.
Interestingly, we find that the central data both of NGC~253 and M82 show an even flatter radio spectrum in comparison to the total spectrum, with particularly strong free-free absorption at small frequencies. This is convincingly explained by the higher electron density in central regions of galaxies, as it is also manifested within our model of NGC~253. 
It also becomes clear from the observations of M82, that the flattening of the radio spectra is a feature that is predominantly originating from the central region of the galaxy. The observations by \cite{2013Adebahr_M82} of the central $\sim$450~pc of M82 clearly show a flat radio spectrum, whereas the halo spectrum (i.e.\ the total spectrum minus the central data) is similarly steep in comparison to our radio synchrotron component. Hence, the flat component of the radio spectrum must result from the central region and can be reconciled with the steep radio synchrotron spectra. 

In Fig.~\ref{fig: maps free-free}, we show modeled free-free emission maps at 1~GHz of our M82-like galaxy to illustrate the dependence of the thermal emission on the observed region and also on the viewing angle. Observing a galaxy edge-on leads to a higher column density along the line of sight, which enhances the thermal free-free emission in the mid-plane. Furthermore, the ratios of radio synchrotron to thermal free-free emission at two radio frequencies shown in the middle and right-hand panels of Fig.~\ref{fig: maps free-free} highlight our conclusion drawn from the radio spectra as discussed above: while the synchrotron emission dominates at low frequencies, thermal free-free emission prevails towards higher frequencies. 

Similarly to NGC~253 and M82, the observed spectrum of NGC~2146 is flatter than the radio synchrotron emission of our simulated galaxy. However, adding a thermal free-free component enables us to reproduce the flat observed spectrum at frequencies above $\sim10$~GHz. In the case of NGC~2146, there are no published data of the central region. We chose to show the central spectrum of NGC~2146 from a region with radius of 0.3~kpc, where we find a pronounced flattening of the central spectrum at low frequencies due to free-fee absorption.

We furthermore note that the radio luminosities given in Table~\ref{Table-Galaxies} for our simulations include radio synchrotron emission and thermal free-free emission, whereas the luminosities adapted in our analysis of the FRC are only the non-thermal radio synchrotron luminosities. We determine the contribution of the thermal free-free emission to the total luminosity at 1.4~GHz to 24.7, 6.3 and 5.7 per cent for NGC~253, M82 and NGC~2146, respectively.

\subsection{Can bremsstrahlung or Coulomb losses flatten radio spectra?\label{sec: radio spectra: bremsstrahlung}}

\begin{figure*}
\begin{centering}
\includegraphics[scale=1]{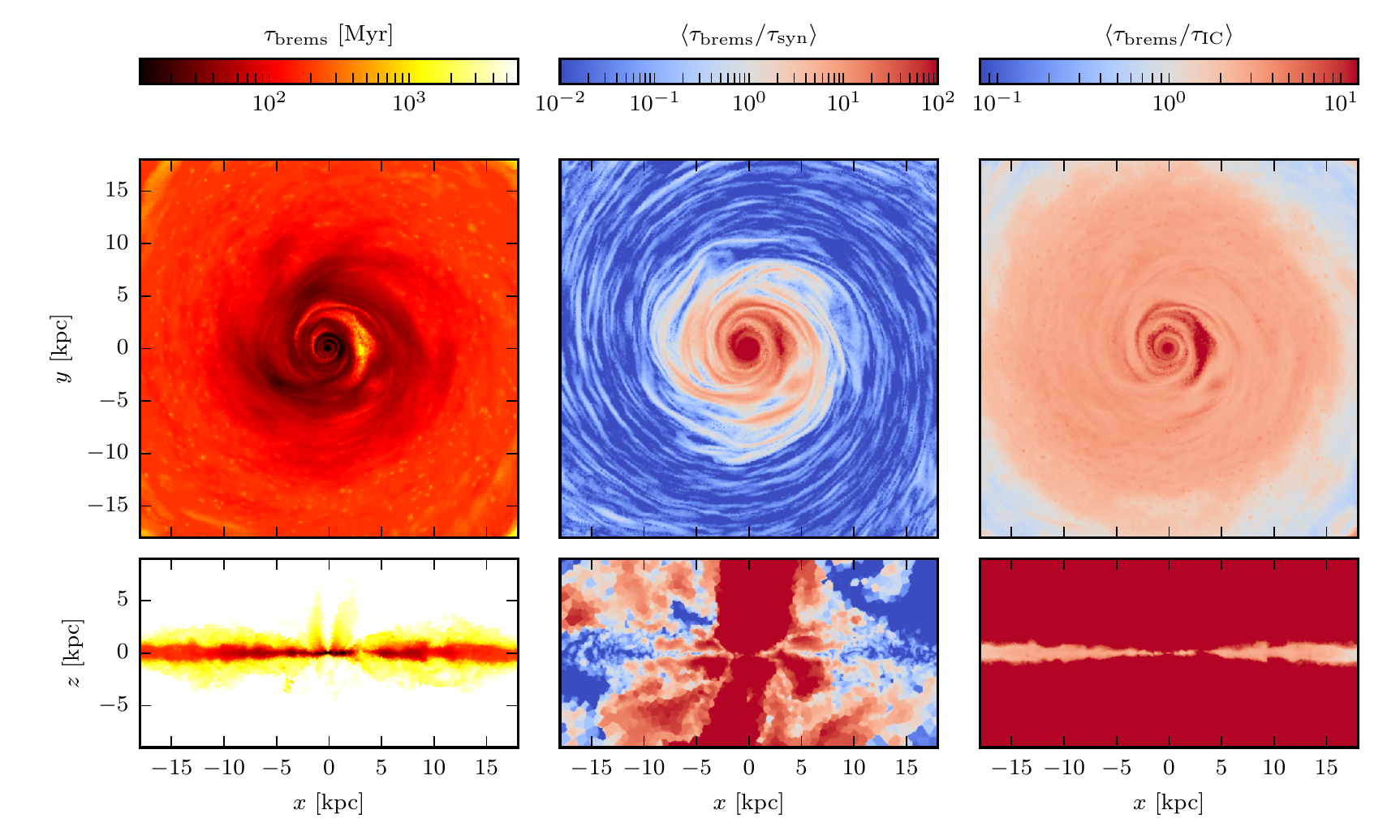}
\par\end{centering}
\caption{Face-on and edge-on maps of the timescale of bremsstrahlung cooling, $\tau_\rmn{brems}$ (left-hand panels), as well as the ratio of $\tau_\rmn{brems}$ to the cooling timescale due to synchrotron and IC emission (middle and right-hand panels, respectively) at an electron energy of 10~GeV. The maps are shown for our fiducial halo with $M_{200}=10^{12}\,\mathrm{M_\odot}$ at 2.3 Gyr and the ratios are averaged over slices with a thickness of 0.5~kpc. }
\label{fig: maps timescales}
\end{figure*}

\begin{figure*}
\begin{centering}
\includegraphics[scale=1]{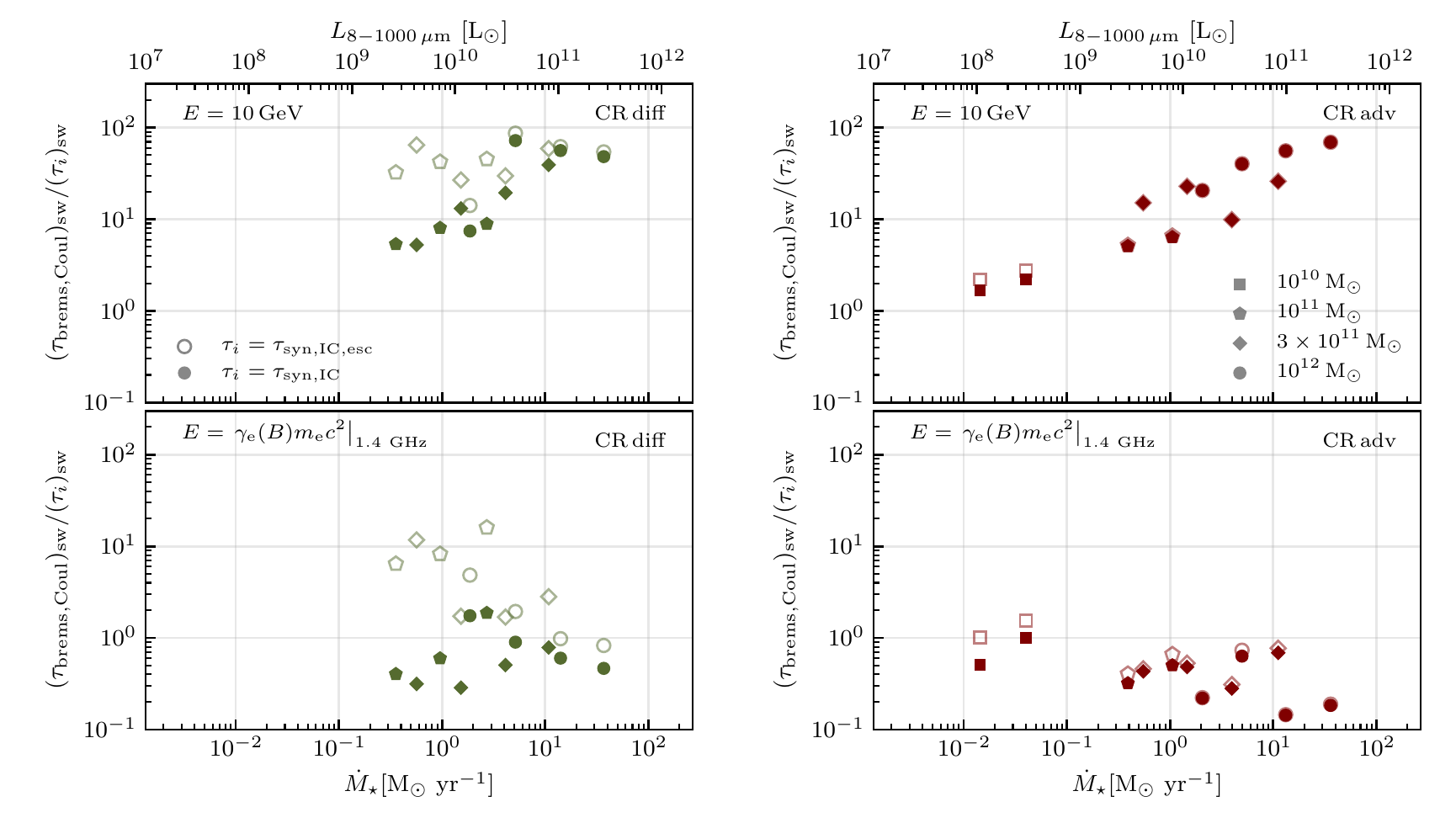}
\par\end{centering}
\caption{Ratios of the synchrotron-weighted timescales of bremsstrahlung and Coulomb cooling $\tau_{\mathrm{brems,Coul}}$ (Eq.~\ref{eq: tau_bremsCoul}) to the timescale of synchrotron and IC cooling $\tau_{\mathrm{syn,IC}}$ (as well as to that of radiative cooling and escape losses $\tau_\mathrm{syn,IC,esc}$, open symbols; see Eqs.~\ref{eq: tau_ICsynchr} and \ref{eq: tau_ICsynchrEscape}) for our `CR diff' model (left-hand panels) and `CR adv' model (right-hand panels). The timescales of each snapshot are calculated by averaging over the corresponding cooling rates (i.e.\ $\tau^{-1}$) from all cells while weighting them with the synchrotron luminosity. In the upper panels, the timescales are computed at an energy of 10~GeV whereas in the lower panels, we adapt the typical electron energy responsible for radio synchrotron emission at $\nu=1.4\,\mathrm{GHz}$, given the magnetic field in each cell (see Eq.~\ref{eq: nu_synchr}). }
\label{fig: ratios_timescsales_sw}
\end{figure*}

In order to assess the possibility that bremsstrahlung cooling flattens radio spectra, we first show in Fig.~\ref{fig: maps timescales} a slice through the disc of the bremsstrahlung cooling timescale at an energy of 10~GeV, which is typically acting on timescales of $50-200$~Myr and is shortest in high-density regions (see Eq.~\ref{eq: tau_brems}). However, synchrotron cooling is faster in the central few kpc and in the outflow, where the magnetic field is strong. In the regions where bremsstrahlung losses are faster than synchrotron cooling, IC cooling kicks in (right-hand panel of Fig.~\ref{fig: maps timescales}). Therefore, bremsstrahlung losses are subdominant in most regions of this galaxy and are not able to significantly shape the electron spectra at the considered electron energy. Consequently, they cannot be responsible for flattening these at the corresponding frequencies. Instead, IC and synchrotron losses steepen the injected electron spectra by unity, i.e., $\alpha_\rmn{e}\approx\alpha_\rmn{inj}+1=3.2$ and hence, we expect $\alpha_{\nu}=(\alpha_\rmn{e}-1)/2=1.1$ in the limit, where all losses are due to IC and synchrotron emission. 

However, as can be seen from the radio spectral index shown in the bottom panels of Fig.~\ref{fig: radio-spectra NGC 253 and M82} and \ref{fig: radio-spectrum NGC2146}, $\alpha_\nu$ only approaches this limit of very high frequencies $\nu\gtrsim10$~GHz. One reason for this is the fact that we include diffusive losses in our steady state equation\footnote{The influence of advection losses in outflows is discussed in Section~\ref{sec: radio spectra: outflows}.}, which only steepen the spectra by $\delta=0.3$ (or 0.5). This implies that we obtain radio spectral indices ranging from 0.75 (or 0.85) if diffusion losses dominate, to 1.1 if IC or synchrotron losses prevail. Furthermore, in regions with particularly large magnetic field strengths, we observe lower-energetic electrons (due to the $\nu_\mathrm{c}$-effect), which are more affected by Coulomb and bremsstrahlung cooling in comparison to  high-energy electrons, where IC and synchrotron losses steepen the spectra.

To quantify this effect and to assess the importance of bremsstrahlung cooling in flattening radio spectra for galaxies along the SFR sequence, we show in Fig.~\ref{fig: ratios_timescsales_sw} the ratio of the combined timescale of bremsstrahlung and Coulomb cooling 
\begin{align}
\tau_{\mathrm{brems,Coul}}= (\tau_{\mathrm{brems}}^{-1} +  \tau_\mathrm{Coul}^{-1})^{-1}
\label{eq: tau_bremsCoul}
\end{align}
to the combined synchrotron and IC timescale, i.e.,
\begin{align}
\tau_{\mathrm{syn,IC}}= (\tau_{\mathrm{syn}}^{-1} +  \tau_\mathrm{IC}^{-1})^{-1},
\label{eq: tau_ICsynchr}
\end{align}
as well as to the combined timescale of synchrotron, IC and escape losses,
\begin{align}
\tau_{\mathrm{syn,IC,esc}}= (\tau_{\mathrm{syn}}^{-1} +  \tau_\mathrm{IC}^{-1} + \tau_\mathrm{esc}^{-1})^{-1}
\label{eq: tau_ICsynchrEscape}
\end{align}
In order to show the cooling timescale ratios in regions that are relevant for synchrotron emission, we weight the corresponding cooling rates ($\tau^{-1}$) with the synchrotron luminosity of each cell when averaging them over the whole galaxy. In both, our `CR diff' and `CR adv' models, we find cooling via bremsstrahlung and Coulomb losses to be slower than IC and synchrotron cooling for electron with energies of $E_\mathrm{e}=10$~GeV, i.e.\ $\tau_\mathrm{brems,Coul}(10~\mathrm{GeV})\approx \tau_\mathrm{brems}(10~\mathrm{GeV})> \tau_\mathrm{syn,IC}$, which coincides with our findings from Fig.~\ref{fig: maps timescales} while accounting for the fact that Coulomb losses are also subdominant in comparison to bremsstrahlung losses at these high energies.

However, if we consider that synchrotron-emitting electrons at $\nu=1.4$~GHz will attain contributions from different energies, depending on the ambient magnetic field (see Eq.~\ref{eq: nu_synchr}), we find that bremsstrahlung and Coulomb losses are not at all negligible in comparison to IC and synchrotron losses in most of our snapshots, but instead, $\tau_\mathrm{brems,Coul}\lesssim \tau_\mathrm{syn,IC}$ at electron energies $E_\mathrm{e}=\left.\gamma_\mathrm{e}(B)m_\mathrm{e}c^2\right|_{\rmn{1.4~GHz}}$. This is due to the fact that we typically probe lower energetic electrons in high magnetic field strengths. The mean Lorentz factor of electrons emitting at 1.4~GHz averaged over all cells ranges between $\gamma_\mathrm{e}=10^4$ and $2\times 10^4$ for all SFRs, corresponding to $E_\mathrm{e}=5$ to $10$~GeV. By contrast, the average $\gamma_\mathrm{e}$ weighted with the synchrotron luminosity in each cell decreases from $10^4$ at small SFRs down to $10^3$ in our starburst galaxies. Hence, the synchrotron emission at $\nu=1.4$~GHz in starburst galaxies with their high magnetic field strengths probes electrons with energies down to $E_\mathrm{e}=0.5$~GeV.

The additional inclusion of escape losses (open symbols in Fig.~\ref{fig: ratios_timescsales_sw}) that become increasingly more important for decreasing SFRs in our `CR diff' model shows that only highly star-forming galaxies are able to maintain $\tau_\mathrm{brems,Coul}\lesssim \tau_\mathrm{syn,IC,esc}$ at the relevant energies, determined by $\gamma_\mathrm{e}(B)$ (Eq.~\ref{eq: nu_synchr}). This is in accordance with our results from Section~\ref{sec:conspiracy at high densities}, where we found that neglecting bremsstrahlung and Coulomb losses increases the radio luminosity at 1.4~GHz for an increasing SFR.
In the `CR adv' model, we observe a similar trend in comparison to the `CR diff' model: at high electron energies (e.g., at 10~GeV), bremsstrahlung and Coulomb losses are subdominant while they become relevant if we account for the $\nu_\mathrm{c}$-effect, where we typically probe lower energetic electrons. If we additionally include escape losses (i.e., only advection losses in this model), they do not significantly influence the considered timescale ratios. 

In summary, we conclude that the synchrotron emission at 1.4~GHz in highly star-forming galaxies probes electrons with $\gamma_\mathrm{e}\sim 10^3$ which are mostly affected by bremsstrahlung and Coulomb losses. By contrast, electrons at higher energies are dominated by synchrotron and IC cooling and consequently, at higher frequencies, the radio synchrotron spectra are steepened by the those cooling processes. Only the modeling of a thermal free-free component as discussed in Section~\ref{sec: radio spectra: thermal and non-thermal radio emission} is thus able to flatten the radio spectra towards high frequencies in starbursts.
We furthermore find an increasing relevance of energy dependent diffusion losses of CR electrons with decreasing SFRs. At the same time, observations at 1.4~GHz probe higher energetic electrons ($\gamma_\mathrm{e}\sim10^4$) at low SFRs so that bremsstrahlung and Coulomb losses do not strongly affect the radio synchrotron luminosity towards low SFRs (see Fig.~\ref{fig: delta_q}).

\subsection{Can outflows in projection flatten radio spectra?}\label{sec: radio spectra: outflows}

\begin{figure*}
\begin{centering}
\includegraphics[scale=1]{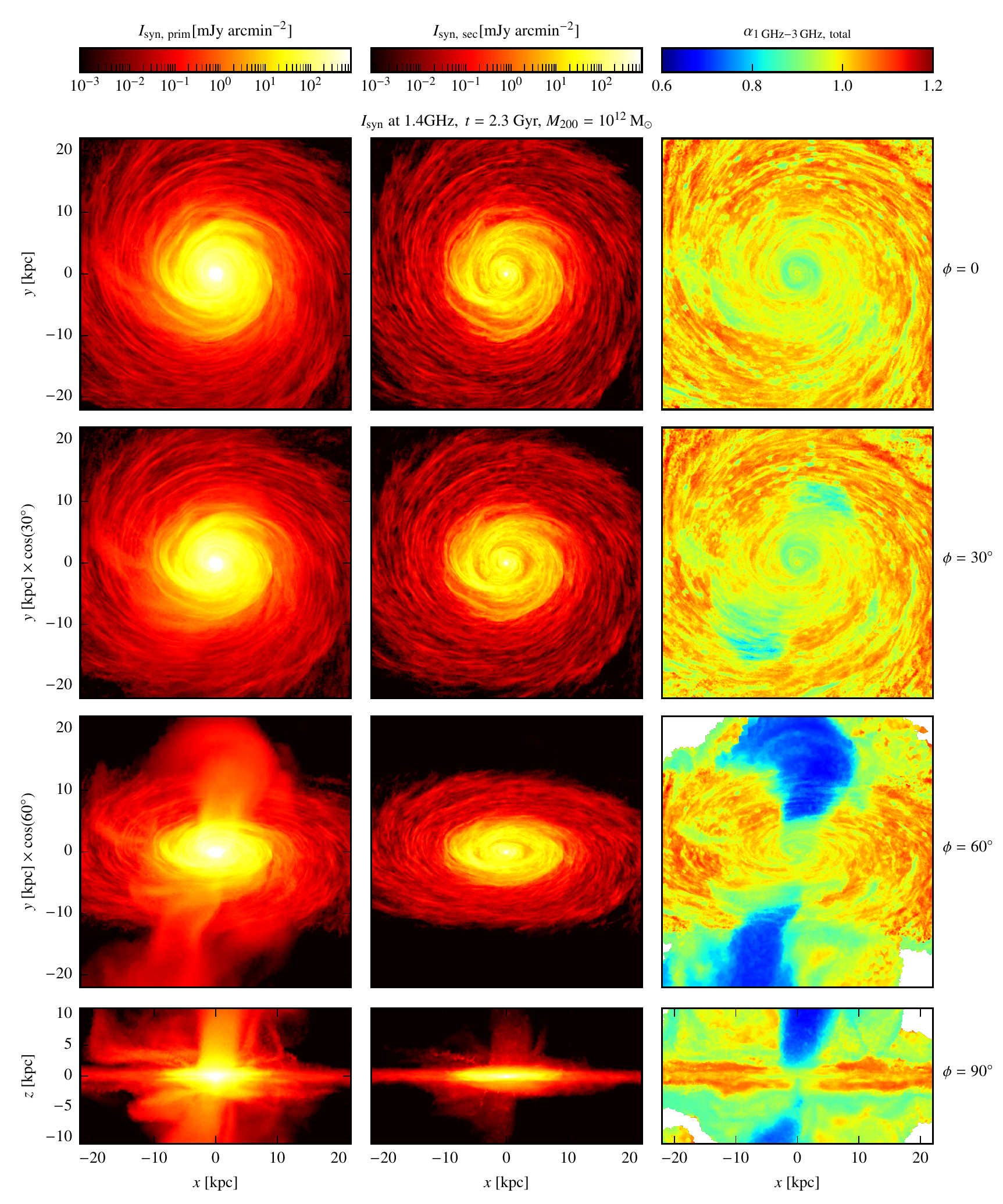}
\par\end{centering}
\caption{From top to bottom, we show projected maps of the radio synchrotron intensity (Eq.~\ref{eq:I_nu definition}) arising from primary (left-hand panels) and secondary electrons (middle panels) observed from different inclination angles (0, 30, 60 and 90 degrees) for our fiducial galaxy with $M_{200}=10^{12}\,\mathrm{M_\odot}$ at $t=2.3\,\mathrm{Gyr}$. The right hand panels show the corresponding maps of the spectral index of the total radio synchrotron emission (i.e.\ the primary plus secondary contribution) between frequencies of 1~GHz and 3~GHz.}
\label{fig: synchr_alpha_maps}
\end{figure*}

We show in Fig.~\ref{fig: synchr_alpha_maps} maps of the projected synchrotron emissivity $I_{1.4\,\mathrm{GHz}}$ resulting from primary (left-hand panels) and secondary electrons (middle panels) for our fiducial halo with $M_{200}=10^{12}\,\mathrm{M_\odot}$ at $t=2.3\,\mathrm{Gyr}$. Because the simulation shown here includes anisotropic diffusion, the radio luminosity from primary electrons dominates over the secondary contribution (as discussed in Section~\ref{sec:secondary vs. primary synchr. emission}). 
From top to bottom, we show projected views of the simulated galaxy, changing the galaxy inclination from a face-on to an edge-on view in steps of $30^{\circ}$. The right-hand panels show spectral index maps of the total synchrotron emission between 1 and 3 GHz.

As discussed above, the spectral index of the steady-state electron population remains unchanged, if advection is the dominant loss process, i.e.\ $\alpha_{\mathrm{e}}=\alpha_{\mathrm{inj}}=2.2$. Hence, the radio spectral index is $\alpha_\nu = (\alpha_{\mathrm{e}}-1)/2=0.6$ in the outflows, where CRs predominantly are advected with the gas into the halo. 
By contrast, the observed spectral index maps of the radio emission of starburst galaxies like M82 \citep{2013Adebahr_M82}, show flat spectral indices mainly in their central regions. We conclude that CR advection in combination with galaxy inclination cannot be responsible for the observed flat radio spectral indices because regions with flat spectral indices (due to advection losses) are seen off-center and only for inclinations larger than about $45^{\circ}$.

\section{Discussion and conclusions}\label{sec: conclusions}

In this work, we model for the first time the radio emission of galaxies in three-dimensional MHD simulations, which evolve the energy density of CR protons self-consistently. To this end, we determine the steady-state spectra of CR protons, primary and secondary electrons in post-processing and assess the relative importance of primaries versus secondaries for models with and without anisotropic CR diffusion. The detailed modeling of the synchrotron and free-free radio emission of star-forming galaxies enables new insight into the underlying physics of the FRC over a broad range of SFRs. In particular, this novel approach sheds light on the long-standing puzzle whether the almost linear FRC implies electron calorimetry and how this compares to the observed hard radio spectra.

While in simulations that only account for CR advection (`CR adv') the secondary electron population is responsible for most of the radio synchrotron emission and independent of SFR, we find in our model that additionally includes anisotropic CR diffusion (`CR diff') that primary electrons dominate the total radio luminosity at 1.4~GHz (see Fig.~\ref{fig: FIR-Radio}). The main reason for this difference is the steepening of the steady-state CR proton spectra due to energy-dependent diffusion, which modifies the shape of the secondary electron source function. Thus, in steady state, the secondary CR electron spectra differ from the primary electron spectra (see Fig.~\ref{fig: CR-spectra}). As a result, due to their steeper spectra, secondary electrons are more affected by the $\nu_\mathrm{c}$-effect, which means that the observed radio synchrotron emission at a given frequency probes lower electron energies for larger magnetic field strengths that are realised in highly star-forming galaxies. This leads to an increasing contribution of secondary emission to the total radio luminosity with SFR in the `CR diff' model, in contrast to the `CR adv' model, where this trend is absent.

To test electron calorimetry, we determine the fraction of available injected CR electron luminosity that is actually converted to synchrotron emission. This calorimetric synchrotron fraction $\eta_\rmn{syn}$ varies from 0.03 to 0.67 among all our simulations with an increasing trend towards larger SFRs (see Fig.~\ref{fig: calorimetric_fraction}). The galaxies in the `CR adv' model show overall larger calorimetric fractions with a weaker dependence on SFR in comparison to the `CR diff' model, where diffusive losses and a weaker saturated magnetic field strength in dwarfs \citep[in comparison to the thermal energy density,][]{2021Pfrommer} decrease $\eta_\rmn{syn}$ towards smaller SFRs. We anticipate that an improved ISM model \citep[e.g.,][]{2021Rathjen} in combination with an improved two-moment CR transport scheme \citep[e.g.,][]{2019ThomasPfrommer} within a cosmological setting that exhibits several epochs of accretion-driven star-forming phases can further amplify the magnetic field strength in these systems via star-formation and accretion-driven turbulence as well as modulate the influence of CR transport on the magnetic dynamo amplification via a more consistent self-generated CR diffusion coefficient. This is expected to weaken the dependence of $\eta_\rmn{syn}$ on SFR. Generally, we find that CR electrons with energies $\gtrsim10$~GeV lose most of their energy through IC interactions with starlight and CMB photons in comparison to synchrotron losses and thus are approximately in the calorimetric limit. However, we also find that the contribution to IC cooling is on average largely independent of SFR at fixed observed radio frequency and hence, enables to use the synchrotron emission as a quasi-calorimetric measure of the SFR (see Fig.~\ref{fig: delta_q}).

Furthermore, we show that the increasing secondary contribution to the total radio luminosity in starbursts is balanced by increasing bremsstrahlung and Coulomb losses, that diminish the radio synchrotron luminosity with increasing densities in those systems (see Fig.~\ref{fig: delta_q}). This finding is in accordance with one-zone models by \citet{2010Lacki}. However in our model, the effect is not as strong: \citet{2010Lacki} find an increase in the logarithm of the TIR-to-radio luminosity ratio $q$ of 0.5 to 0.6 if secondaries are not included in starbursts, as well as a corresponding decrease if secondaries are included but non-synchrotron losses are disregarded. By contrast, our `CR diff' model predicts a change of $q$ of about 0.2 in our highest star-forming galaxy when we neglect the secondary radio emission. If non-synchrotron cooling is disregarded, $q$ changes by 0.2 to 0.4.
Note that we refrain from analysing this effect as a function of gas surface density, because this observable strongly depends on the viewing angle of the galaxy, whereas the SFR is a more robust global quantity. 

Finally, we examine three possible solutions for the problem of the discrepancy between the observed hard radio spectra and the theoretically expected steep radio synchrotron spectra in the calorimetric limit, i.e.\ that CR electrons lose their energy primarily to radiative (synchrotron and IC) processes before they can escape from the galaxies.
\begin{enumerate}
\item Bremsstrahlung and Coulomb cooling processes are only able to flatten low-energy electron spectra, and hence, low-frequency radio spectra. However, these processes only play a minor role in the observed spectral flattening because (diffusive) escape losses are more important, especially at lower SFRs  (see Fig.~\ref{fig: ratios_timescsales_sw}). 
\item We also find that advection losses cannot be primarily responsible for the observed low spectral indices because they only generate hard CR electron spectra in outflows. This implies hard radio spectra off-center, which is in direct conflict with the observed hard \textit{central} radio spectra (see Fig.~\ref{fig: synchr_alpha_maps}).
\item Our preferred solution is thermal free-free emission that starts to dominate the total radio spectrum at frequencies of several to tens of GHz and hardens the observed radio spectrum (see Figs.~\ref{fig: radio-spectra NGC 253 and M82} and \ref{fig: radio-spectrum NGC2146}). Thermal free-free emission is not only able to harden the radio spectra of our simulated galaxies similar to NGC~253, M82 and NGC~2146 at high frequencies, but the involved absorption process, i.e.\ free-free absorption, coincides with the spectral hardening of the central spectrum of NGC~253. 
\end{enumerate}

This interpretation of thermal free-free emission being the main driver for hardening the radio spectra at high frequencies is corroborated by the observed spectrum of the central $\sim450$~pc of M82: subtracting this central emission region from the total spectrum yields a spectrum that is as steep as the radio synchrotron spectrum resulting from CR electrons close to the fully cooled limit, with spectral indices up to 1.1. Consequently, the hardening of the spectra at high frequencies must mainly originate from the central region of starbursts and can be reconciled with steep radio synchrotron spectra, that dominate the emission outside the dense, central starburst region and the total spectrum at lower frequencies, i.e., below $\sim$3, 10 or 20~GHz in our modeling of NGC~253, NGC~2146 and M82, respectively. 

These insights into the radio emission from galaxies and the physics of the FRC necessarily require a spectral modelling of the different CR populations in full three-dimensional MHD-CR simulations, which demonstrates the power of this approach. While we confirm several findings of one-zone models \citep[e.g.,][]{2010Lacki}, our approach exhibits less free parameters and thus can be considered to be more predictive. Most importantly, because these simulations evolve the magnetic field and the CR energy density, we can use this modeling to link the simulated radio emission to observational data to quantify the effect of CR feedback on galaxy formation.

Clearly, the presented radio analysis of steady-state CR spectra need to be complemented by full spectral-dynamical simulations of CR protons \citep{2020MNRAS.491..993G} and electrons \citep{2019MNRAS.488.2235W, Winner2020, 2020arXiv200906941O}. Moreover, the employed one-moment approach for CR transport \citep{2017aPfrommer} will be improved with a two-moment CR hydrodynamics model that is coupled to the Alfv\'en wave energy density, which delivers a realistic (spatially and temporally varying) CR diffusion coefficient in the self-confinement picture \citep{2019ThomasPfrommer,2021arXiv210508090T,2021MNRAS.503.2242T}. Finally, pursuing MHD-CR simulations of galaxy formation in a cosmological setting \citep{2020Buck,2021MNRAS.501.4184H} will be required to obtain realistic (bursty) star-formation histories. These have different epochs of turbulent driving, which can potentially amplify the magnetic field in dwarf galaxies furthermore to come into equipartition with the thermal energy density so that it does not saturate at a sub-equipartition level \citep{2021Pfrommer} and reaches the FRC.

\section*{Acknowledgements}
The authors acknowledge support by the European Research Council under ERC-CoG grant CRAGSMAN-646955.

\section*{Data Availability}
The data underlying this article will be shared on reasonable request to the corresponding author. The \arepo code is publicly available.

\bibliographystyle{mnras}
\bibliography{literatur}

\appendix

\section{Radiation processes}
 
\subsection{Synchrotron emission}\label{App: Synchrotron emission}

Each relativistic electron with charge $e$ that is accelerated by an ambient magnetic field $B$ emits synchrotron radiation. The resulting power emitted per unit frequency $\nu$ is given by \citep{1986rpa..book.....R}
\begin{align}
P(\nu, \gamma_{\mathrm{e}} )=\frac{\mathrm{d}E}{\mathrm{d}\nu\mathrm{d}t}=\frac{\sqrt{3}e^{3}B\sin\alpha_{\mathrm{pitch}}}{m_{\mathrm{e}}c^{2}}F(x),
\label{eq: synchr. power per electron}
\end{align}
where $\alpha_\mathrm{pitch}$ is the pitch angle, $x=\nu/\nu_\mathrm{c}$, the critical frequency is $\nu_\mathrm{c}=3/(4\upi)\,\gamma_{\mathrm{e}}^{3}\omega_{B}\sin\alpha_\mathrm{pitch}$, and the frequency of gyration is $\omega_{B}=eB/(\gamma_{\mathrm{e}} m_{e}c)$.
Furthermore, the dimensionless synchrotron kernel $F(x)$ is defined as
\begin{align}
F(x)=x\intop_{x}^{\infty}K_{5/3}(\xi)\mathrm{d}\xi,
\label{eq: F(x)}
\end{align}
where $K_{5/3}$ is the modified Bessel function of order 5/3. 
For a population of electrons with a distribution $f_{\mathrm{e}}(p)$, the total emissivity, i.e., the emitted energy per unit time, volume and frequency, is obtained by integrating over the electron distribution:
\begin{align}
j_{\nu}(\nu)= E\frac{\mathrm{d}N_{\gamma}}{\mathrm{d}\nu \mathrm{d}V \mathrm{d}t}=\frac{\sqrt{3}e^{3}B_{\perp}}{m_{e}c^{2}}\intop_{0}^{\infty}f_{\mathrm{e}}(p)F(x)\mathrm{d}p,
\label{eq: Synchr. total synchrotron emissivity_2}
\end{align}
where $B_\perp$ is the component of the magnetic field perpendicular to the line of sight and the argument of the synchrotron kernel depends on the electron momentum via the dependence of the critical frequency on the Lorentz factor, $\gamma_\rmn{e}=\sqrt{1+p^2}$. Since the integration over the modified Bessel function is numerically expensive, we use an analytical approximation by \citet{2010PhRvD..82d3002A}, which reads
\begin{align}
\widetilde{F}(x) &\approx2.15x^{1/3}(1+3.06x)^{1/6} \nonumber \\
& \times \frac{1+0.884x^{2/3}+0.471x^{4/3}}{1+1.64x^{2/3}+0.974x^{4/3}}e^{-x}.
\label{eq: synchr. aharonian F(x)}
\end{align}
This function peaks at $x=0.2858$, but its expectation value lies at $x=2.13$. 

\subsection{Thermal free-free emission and absorption}\label{App: free-free-emission}

Thermal electrons are deflected in the Coulomb field of ions of charge $Ze$ and emit thermal free-free emission. The resulting emissivity from a medium with electron density $n_\mathrm{e}$, ion density $n_\mathrm{i}$ and temperature $T$ is given by \citep{1986rpa..book.....R}
\begin{align}
j_{\nu,\mathrm{ff}} = ~~&6.8 \times 10^{-38}~\mathrm{erg\,s^{-1}\,cm^{-3}\,Hz^{-1}}\nonumber\\
&\times Z^2 \frac{n_{\mathrm{e}}\,n_{\mathrm{i}}}{\mathrm{cm^{-6}}} \, T_1^{-0.5}\, e^{-h\nu/(k_{\mathrm{B}} T)}\, \overline{g}_{\mathrm{ff}},
\end{align}
where $T_1 = T/(1\,\mathrm{K})$, $h$ is Planck's constant and $k_\mathrm{B}$ denotes Boltzmann's constant.
The corresponding absorption coefficient of free-free absorption in the Rayleigh-Jeans regime, i.e.\ for $h\nu \ll k_{\mathrm{B}}T $, reads \citep{1986rpa..book.....R}
\begin{align}
\kappa_{\mathrm{ff}}(\nu) = 0.018\,T_1^{-1.5}\,Z^2 \frac{n_{\mathrm{e}}\,n_{\mathrm{i}}}{\mathrm{cm^{-6}}} \, \left( \frac{\nu}{\mathrm{Hz}} \right)^{-2}\, \overline{g}_{\mathrm{ff}} \, \mathrm{cm^{-1}}.
\label{eq: kappa ff}
\end{align}
Here, the mean Gaunt factor $\overline{g}_\mathrm{ff}$ in the ``small-angle, classical region'' is given by \citep{1973NovikovThorne}
\begin{align}
\overline{g}_{\mathrm{ff}}  = \frac{\sqrt{3}}{\upi} \ln \left[ \frac{1}{4\xi^{5/2}Z} \left(\frac{k_{\mathrm{B}}T}{h\nu} \right)   \left( \frac{k_{\mathrm{B}}T}{13.6\,\mathrm{eV}} \right)^{0.5} \right],
\end{align}
where $\xi=\exp(\gamma_\rmn{E})\approx 1.781$ and $\gamma_\rmn{E}$ denotes Euler's constant.
We assume a temperature of $T=8000~$K for the warm ionized medium and approximate the electron density $n_\mathrm{e}$ by taking a constant fraction $\xi_\mathrm{e}$ of the electron density provided in our pressurised ISM model \citep{2003SpringelHernquist} to parametrize the effect of radiation feedback in the dense star-forming/bursting regions and further assume $n_{\mathrm{e}}\approx n_{\mathrm{i}}$ and $Z=1$. 
The optical depth due to free-free absorption is calculated via the integral of the absorption coefficient along the line of sight, i.e.
\begin{align}
\tau_{\mathrm{ff}}(\nu,s_0,s_1)& =\intop_{s_0}^{s_1} \kappa_{\mathrm{ff}} (\nu)\mathrm{d}s\\
&=  0.018\,T_1^{-1.5}\,Z^2  \left( \frac{\nu}{\mathrm{Hz}} \right)^{-2}\, \overline{g}_{\mathrm{ff}}  \intop_{s_0}^{s_1} \left(\frac{ n_{\mathrm{e}} }{\mathrm{cm^{-3}}}\right)^2 \mathrm{d}s,
\label{eq: tau ff}
\end{align}
where $\mathrm{d}s=\sin\phi\, \mathrm{d}y + \cos\phi\,\mathrm{d}z$ and $\phi$ denotes the inclination angle.
Because for a constant temperature, $j_{\nu,\mathrm{ff}}/\kappa_{\mathrm{ff}}$ is constant along the line of sight and the solution of the radiative transfer equation \citep[e.g.][]{1986rpa..book.....R} is straight forward and yields an observed intensity of free-free emission
\begin{align}
4 \upi\, I_{\nu,\mathrm{ff}}(s_1) = \frac{j_{\nu,\mathrm{ff}}}{\kappa_{\mathrm{ff}}} \left\{ 1-\exp [-\tau_{\mathrm{ff}}(\nu,s_0,s_1)] \right\}.
\label{eq: I_nu free-free-emission}
\end{align}

Also, the emitted synchrotron spectrum is affected by free-free absorption, which is typically relevant at low frequencies, as well as synchrotron self-absorption (SSA). The absorption coefficient and the corresponding optical depth of SSA reads \citep{1986rpa..book.....R}
\begin{align}
\kappa_{\mathrm{SSA}} (\nu)& = -\frac{c^2}{8\upi\nu^2} \intop_0^{\infty}  \mathrm{d}p_{\mathrm{e}} P(\nu, p_{\mathrm{e}} ) \left(1+p_{\mathrm{e}}^2 \right)  \frac{\partial}{\partial p_{\mathrm{e}} }\left[\frac{f_{\mathrm{e}} (p_{\mathrm{e}} )}{\beta \left(1+p_{\mathrm{e}} ^2\right)}\right]
\label{eq: kappa SSA}
\end{align}
and
\begin{align}
\tau_{\mathrm{SSA}}(\nu,s_0,s_1) &=\intop_{s_0}^{s_1} \kappa_{\mathrm{SSA}} (\nu)\mathrm{d}s.
\label{eq: tau SSA}
\end{align}
In order to calculate the absorbed synchrotron spectrum, we use the formal solution of the radiative transfer equation for an emissivity $j_\nu$ and the optical depth $\tau=\tau_{\mathrm{SSA}}+\tau_\mathrm{ff}$., i.e.
\begin{align}
4 \upi\, I_{\nu}(s_1) = \intop_{s_0}^{s_1} j_{\nu}(s) \exp[-\tau(\nu, s,s_1)] \mathrm{d}s.
\label{eq: I_nu rad. transfer}
\end{align}
In practice, we solve Eq.~(\ref{eq: I_nu rad. transfer}) by computing $j_{\nu}$ on slices that are oriented perpendicular to the line of sight $s$ and equidistantly spaced along $s$ through the simulation cube and cumulatively add the optical depth. If we observe a simulated galaxy edge-on, for instance, $\mathrm{d}s=\mathrm{d}y$ and we obtain two-dimensional slices of the optical depth and the synchrotron emissivity in the $x-z$ plane, i.e.\ $\tau (x,z)$ and $j_\nu(x,z)$. This results in a two-dimensional projection of the intensity $I_\nu(x,z)$ after adding up the slices from the front to the back of the simulation. 
This procedure also allows us to construct the absorbed spectrum from a different viewing angle by initially rotating the cube around the $x$-axis, which yields $I_\nu(\mathbfit{r}_\perp) $, where $\mathbfit{r}_\perp$ is the vector in the plane perpendicular to the line of sight.
The observed flux density from an object emitting at a luminosity distance $d$ is then obtained by integrating over the area $A$ and solid angle $\Omega$
 \begin{align}
F_\nu = \frac{1}{4 \upi d^2} \int_\Omega \int_A I_\nu(\mathbfit{r}_\perp) \mathrm{d}^2r_\perp \mathrm{d}\Omega.
\label{eq: F_nu}
\end{align}

\subsection{Radio spectra for different central radii}\label{app: R_central}

\begin{figure}
\includegraphics[scale=1]{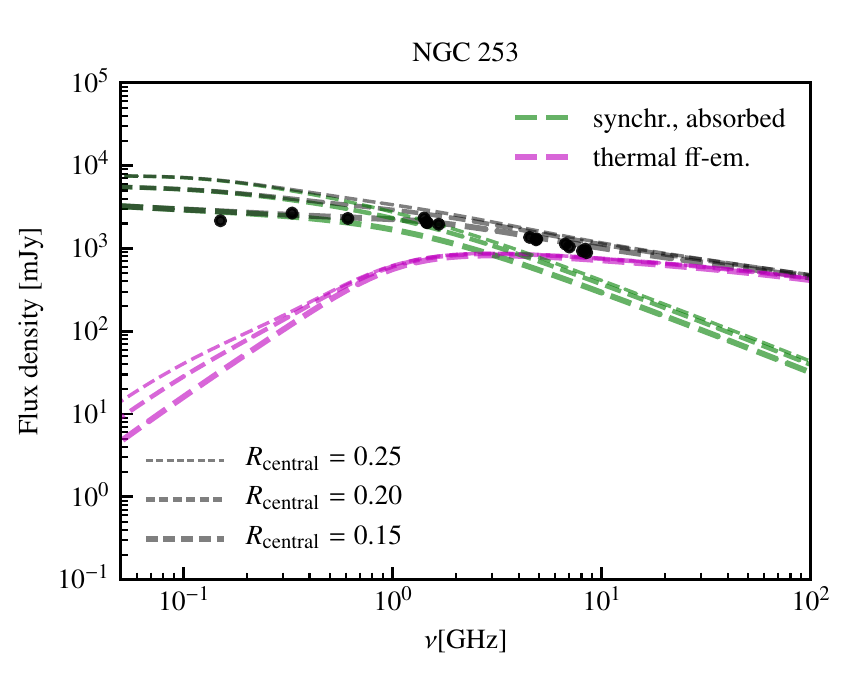}
\caption{We compare the central radio spectrum of our simulated NGC~253-like galaxy (see Table~\ref{Table-Galaxies}) to observational data by \citet{2017Kapinska} of the central region. We compare the spectra within three different radial regions (as indicated in the legend) and show the resulting thermal free-free emission (magenta) and radio synchrotron spectrum (green).}
\label{fig: radio_spectrum_NGC253_Rcuts}
\end{figure}

The radio spectra of the central regions of our analysed galaxies shown in the middle panels of Fig.~\ref{fig: radio-spectra NGC 253 and M82} and \ref{fig: radio-spectrum NGC2146} are calculated by adapting the radii summarised in Table~\ref{Table-Galaxies}. Here, we analyse the effect of choosing different central radii on the spectral shape of free-free-absorbed synchrotron spectra and the spectra of free-free emission. As an example, we show the central spectrum of NGC~253. Decreasing the radius of the central region of the galaxy yields a lower synchrotron flux, which additionally suffers from more absorption losses due to the higher central gas densities, leading to a stronger turn-over at low frequencies.

\begin{figure*}
\begin{centering}
\includegraphics[scale=1]{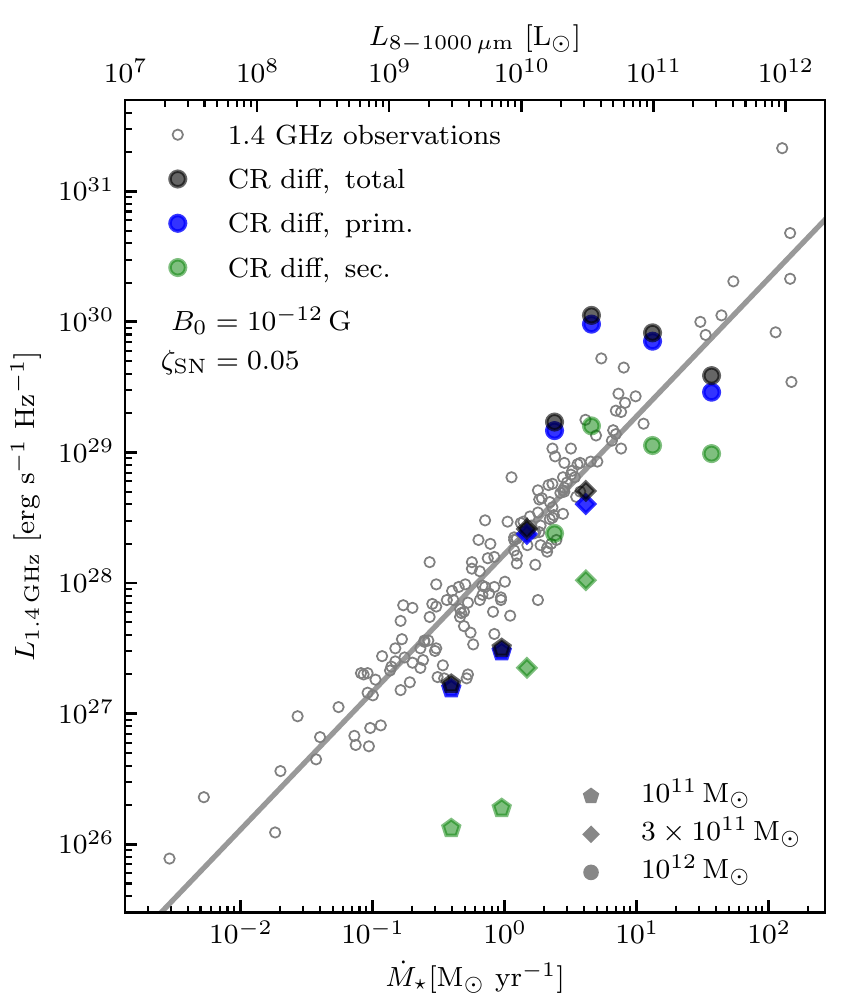}\includegraphics[scale=1]{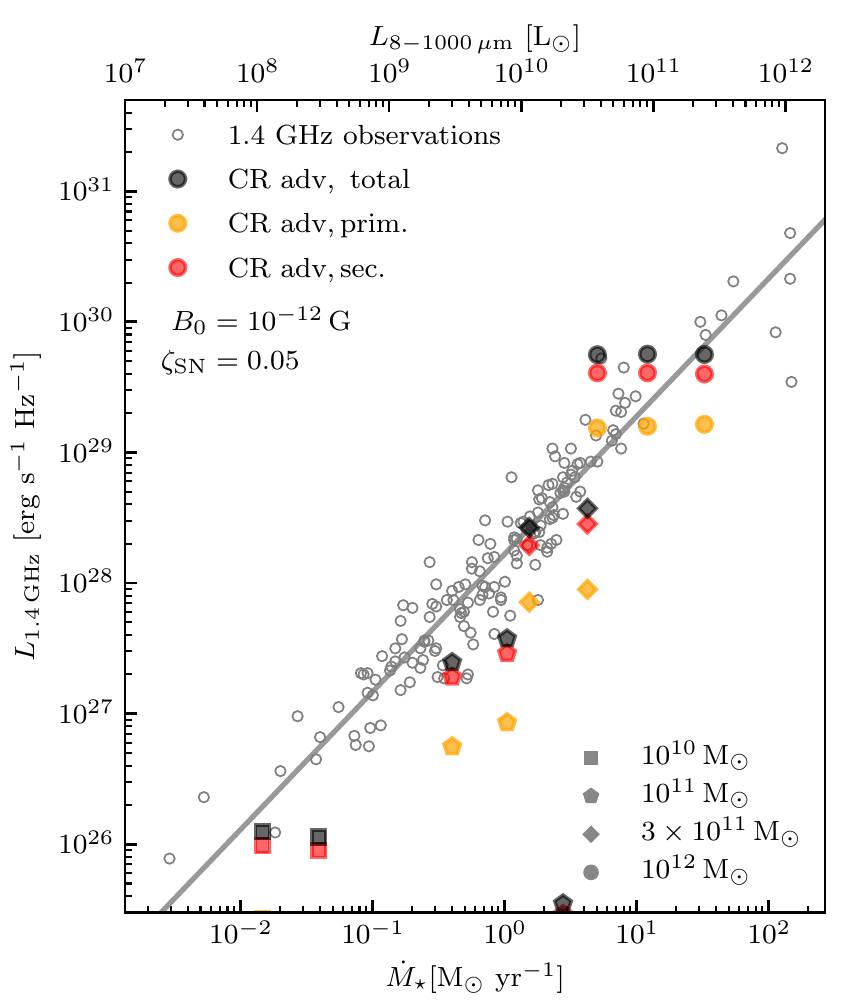}
\includegraphics[scale=1]{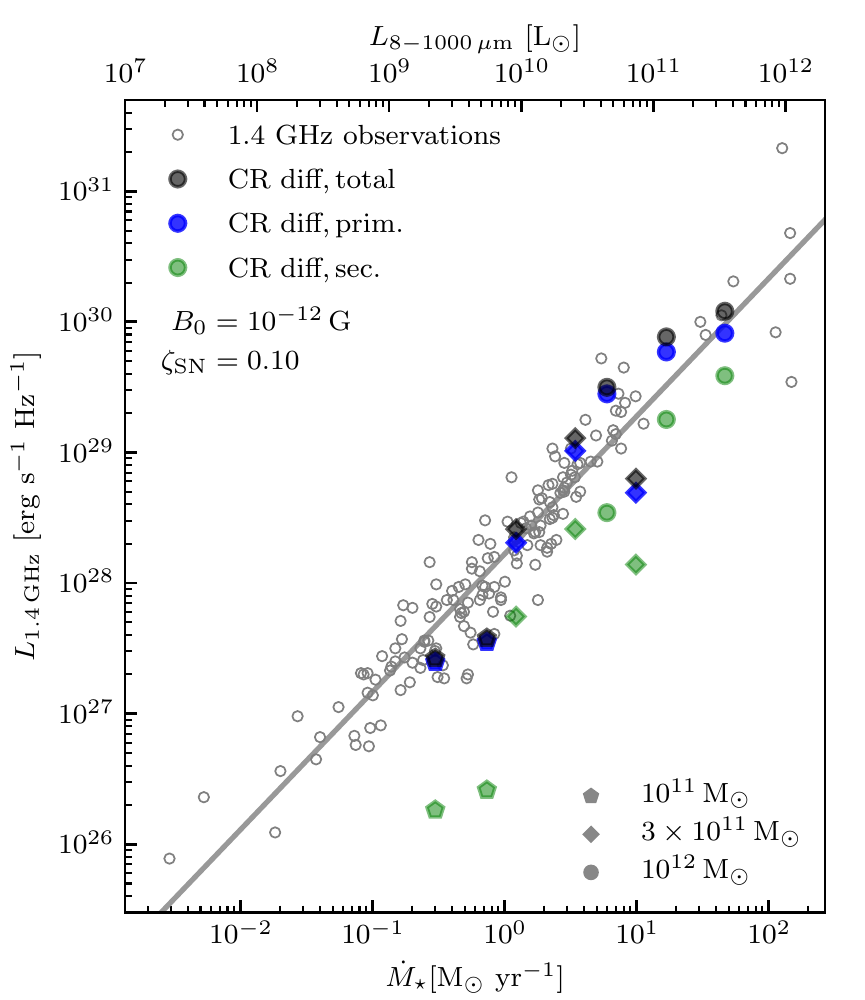}\includegraphics[scale=1]{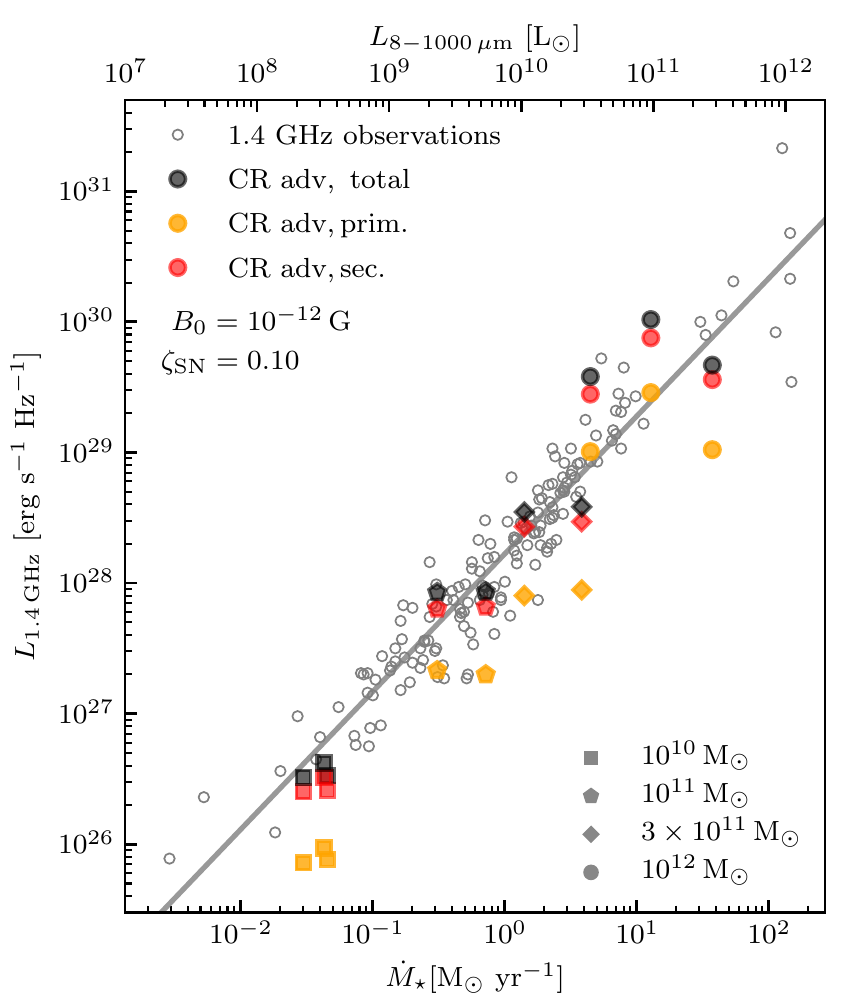}
\par\end{centering}
\caption{Parameter variation of the FRC. In contrast to Fig.~\ref{fig: FIR-Radio}, here we adopt a lower initial magnetic field strength $B_0 = 10^{-12}\,\mathrm{G}$ (top panels). In the bottom panels, we additionally increase the CR injection efficiency by a factor of two, $\zeta_{\mathrm{SN}}=0.1$. Generally, our radio luminosities match the observed scatter. }
\label{fig: FIR-Radio-variations}
\end{figure*}

\section{Parameter variation of the FRC}\label{app: model variation FRC}

In Fig.~\ref{fig: FIR-Radio-variations}, we show the FRC with variations of the parameters of the model discussed in Section~\ref{Sec: FIR-Radio}, where we adapt $\zeta_\mathrm{SN}=0.05$ and $B_0=10^{-10}\,\mathrm{G}$ (see Fig.~\ref{fig: FIR-Radio}). Instead, the simulations shown in the top panels of Fig.~\ref{fig: FIR-Radio-variations} have an initial magnetic field of $B_0=10^{-12}\,\mathrm{G}$. Still, the observed FRC is reproduced both in our `CR diff' and `CR adv' models, respectively. There is only one outlier in the $10^{12}~\rmn{M}_\odot$ halo mass simulation in the `CR diff' model, that overshoots the FRC beyond the observed scatter, at a SFR of $4.6~\mathrm{M_\odot~yr^{-1}}$. This illustrates the different dynamo actions taking place in the different simulations. As shown in \cite{2021Pfrommer}, after $t\approx 1$~Gyr, the averaged magnetic energy density of a simulation with $M_{200}=10^{12}\,\mathrm{M_\odot}$ initiated with a lower magnetic field of $B_0=10^{-12}$~G manages to overtake the simulations that started with a higher initial magnetic field of $B_0=10^{-10}$~G.  However, for smaller halos, the magnetic field grows at a smaller rate in the simulations with $B_0=10^{-12}$~G \citep{2021Pfrommer} and hence, they tend to fall short of the FRC.

The lower panels of Fig.~\ref{fig: FIR-Radio-variations} show the FRC of the simulation model with $B_0=10^{-12}\,\mathrm{G}$, but where CRs are injected with a higher acceleration efficiency and obtain a fraction of $\zeta_\mathrm{SN}=0.10$ of the SN explosion energy, that is a factor of two larger than in our fiducial model.  The combination of a lower initial magnetic field, but on the other hand a higher acceleration efficiency of CRs at SNe yields very similar results in comparison to our fiducial model and also matches the FRC within the observed scatter. However, we found in \citetalias{2021WerhahnII} that this acceleration efficiency of 10 per cent is inconsistent with the observed FIR-gamma-ray relation, because it over-predicts the gamma-ray luminosity $L_{0.1-100~\mathrm{GeV}}$ in both the `CR diff' and `CR adv' models. This highlights the importance of comparing theoretical models with observations across all accessible electromagnetic frequencies.

\bsp

\label{lastpage}
\end{document}